\documentclass[]{emulateapj}
\usepackage{apjfonts}
\usepackage{graphics,amssymb}
\usepackage{bm}

\shortauthors{Lawrence et al.}
\shorttitle{Coyote III}

\begin{document}
\journalinfo{The Astrophysical Journal}
\submitted{Submitted to the Astrophysical Journal}


  \vbox to 0pt{\vss
                    \hbox to 0pt{\hskip 440pt\rm LA-UR-09-06131\hss}
                   \vskip 25pt}

\title{The Coyote Universe III: Simulation Suite and Precision 
Emulator for the Nonlinear Matter Power Spectrum}
\author{Earl~Lawrence\altaffilmark{1}, Katrin~Heitmann\altaffilmark{2},
        Martin White\altaffilmark{3}, David~Higdon\altaffilmark{1},
        Christian~Wagner\altaffilmark{4,5},
        Salman~Habib\altaffilmark{6},
        and Brian~Williams\altaffilmark{1}}

\affil{$^1$ CCS-6, CCS Division, Los Alamos National Laboratory, Los
Alamos, NM 87545}
\affil{$^2$ ISR-1, ISR Division,  Los
Alamos National Laboratory, Los Alamos, NM 87545}
\affil{$^3$ Departments of Physics and Astronomy, University of
California, Berkeley, CA 94720}
\affil{$^4$ Astrophysikalisches Institut Potsdam (AIP),
An der Sternwarte 16, D-14482 Potsdam}
\affil{$^4$ Institut de Ci\`encies del Cosmos (ICC), Universitat de Barcelona, Marti i 
Franqu\`es 1, E-08028 Barcelona}
\affil{$^6$ T-2, Theoretical Division, Los
Alamos National Laboratory, Los Alamos, NM 87545}

\begin{abstract} 
  Many of the most exciting questions in astrophysics and cosmology,
  including the majority of observational probes of dark energy,
  rely on an understanding of the nonlinear regime of structure
  formation. In order to fully exploit the information available from
  this regime and to extract cosmological constraints, accurate
  theoretical predictions are needed. Currently such predictions can
  only be obtained from costly, precision numerical simulations. This
  paper is the third in a series aimed at constructing an accurate
  calibration of the nonlinear mass power spectrum on Mpc scales for a
  wide range of currently viable cosmological models, including dark
  energy. The first two papers addressed the numerical challenges, and
  the scheme by which an interpolator was built from a carefully
  chosen set of cosmological models. In this paper we introduce the
  ``Coyote Universe'' simulation suite which comprises nearly 1,000
  $N$-body simulations at different force and mass resolutions,
  spanning 38 $w\,$CDM cosmologies. This large simulation suite
  enables us to construct a prediction scheme, or emulator, for the
  nonlinear matter power spectrum accurate at the percent level out to
  $k\simeq 1\,h\,{\rm Mpc}^{-1}$. We describe the construction of the
  emulator, explain the tests performed to ensure its accuracy, and
  discuss how the central ideas may be extended to a wider range of
  cosmological models and applications. A power spectrum emulator code
  is released publicly as part of this paper. 
\end{abstract}


\section{Introduction}

\label{sec:intro}

During the last three decades, cosmology has made tremendous progress:
from order of magnitude estimates to measurements of key cosmological
parameters approaching percent level accuracy. The standard model of
cosmology, based on the growth of structure by gravitational
instability, has been impressively validated on large scales where the
perturbations to the Friedman model are small. While the model is
successful at predicting or reproducing a wide array of observations,
it contains several mysterious elements with perhaps the most
mysterious being the accelerated expansion of the Universe
\citep{riess,perlmutter}. The accelerated expansion may be caused by a
dark energy or hinting at a modification of general relativity on the
largest scales.

To date most of the best known cosmological parameters have been
constrained primarily from the study of anisotropies in the cosmic
microwave background (CMB) radiation. However, large-scale structure
probes play an important role in breaking parameter degeneracies and
in constraining conditions in the late-time Universe. Such probes are
becoming ever more precise, with future surveys aiming at measurements
approaching percent level accuracy to better characterize the Universe
in which we live. Techniques based on the observation and analysis of
cosmic structure include the use of baryon acoustic oscillations,
redshift space distortions, weak lensing measurements and the
abundance of clusters of galaxies; they stand to play a pivotal role
in improving our understanding of the dynamics of the Universe. (For a
recent discussion regarding improvements on dark energy constraints
from combining different probes, see, e.g.~\citealt{DETF,FOM}). The
large scale structure of the Universe contains information about both
the geometry, as well as the dynamics of structure formation. In
combination these two pieces of information can help distinguish
between dark energy or a modification of general relativity as the
prime cause of cosmic acceleration.

On small scales, the cosmological interpretation of structure
formation probes is complicated due to the nonlinear physics involved.
Commonly-used fitting functions for e.g., the power spectrum
\citep{PD96,Smi03} have poorly characterized systematics and are no
longer adequate for precision work. Absent a controlled theoretical
framework, direct use of simulations (augmented with phenomenological
parameters as appropriate) is essential if the physics is to be more
correctly captured. The simulation codes need to be adequately tested
to ensure they meet the new demands being placed upon them. The
simulations which pass these goals are often very expensive and only a
restricted number of runs can be performed. This in turn puts a
premium on developing very efficient strategies to constrain
parameters from limited observations and simulations. High-precision
prediction schemes for different statistics are essential to succeed
in this task.

\begin{table*}
\begin{center} 
\caption{The parameters for the 37+1 models which define the sample
  space; $k_{\rm nl}$ is measured in Mpc$^{-1}$. See text for further details.}
\vspace{-0.3cm}
\begin{tabular}{ccccccccc|ccccccccc}
\# & $\omega_m$ & $\omega_b$ & $n_s$ & $-w$ & $\sigma_8$ & $h$ &
$k_{\rm nl}^{z=0}$ & $k_{\rm nl}^{z=1}$ & 
\# & $\omega_m$ & $\omega_b$ & $n_s$ & $-w$ & $\sigma_8$ & $h$ &
$k_{\rm nl}^{z=0}$ & $k_{\rm nl}^{z=1}$ \\ \hline 
 M000 & 0.1296 & 0.0224 & 0.9700 & 1.000 & 0.8000 & 0.7200 & 0.12 & 0.19 &
M019 & 0.1279 & 0.0232 & 0.8629 & 1.184 & 0.6159 & 0.8120 & 0.15 & 0.24 \\ 
 M001 & 0.1539 & 0.0231 & 0.9468 & 0.816 & 0.8161 & 0.5977 & 0.11 & 0.18 &
M020 & 0.1290 & 0.0220 & 1.0242 & 0.797 & 0.7972 & 0.6442 & 0.11 & 0.18 \\ 
 M002 & 0.1460 & 0.0227 & 0.8952 & 0.758 & 0.8548 & 0.5970 & 0.10 & 0.17 &
M021 & 0.1335 & 0.0221 & 1.0371 & 1.165 & 0.6563 & 0.7601 & 0.16 & 0.25 \\ 
 M003 & 0.1324 & 0.0235 & 0.9984 & 0.874 & 0.8484 & 0.6763 & 0.11 & 0.17 &
M022 & 0.1505 & 0.0225 & 1.0500 & 1.107 & 0.7678 & 0.6736 & 0.13 & 0.22 \\ 
 M004 & 0.1381 & 0.0227 & 0.9339 & 1.087 & 0.7000 & 0.7204 & 0.14 & 0.22 &
M023 & 0.1211 & 0.0220 & 0.9016 & 1.261 & 0.6664 & 0.8694 & 0.15 & 0.23 \\ 
M005 & 0.1358 & 0.0216 & 0.9726 & 1.242 & 0.8226 & 0.7669 & 0.12 & 0.20 &
M024 & 0.1302 & 0.0226 & 0.9532 & 1.300 & 0.6644 & 0.8380 & 0.16 & 0.24 \\ 
M006 & 0.1516 & 0.0229 & 0.9145 & 1.223 & 0.6705 & 0.7040 & 0.14 & 0.24 &
M025 & 0.1494 & 0.0217 & 1.0113 & 0.719 & 0.7398 & 0.5724 & 0.12 & 0.20 \\ 
 M007 & 0.1268 & 0.0223 & 0.9210 & 0.700 & 0.7474 & 0.6189 & 0.11 & 0.18 &
M026 & 0.1347 & 0.0232 & 0.9081 & 0.952 & 0.7995 & 0.6931 & 0.11 & 0.18 \\ 
 M008 & 0.1448 & 0.0223 & 0.9855 & 1.203 & 0.8090 & 0.7218 & 0.12 & 0.20 &
M027 & 0.1369 & 0.0224 & 0.8500 & 0.836 & 0.7111 & 0.6387 & 0.12 & 0.19 \\ 
 M009 & 0.1392 & 0.0234 & 0.9790 & 0.739 & 0.6692 & 0.6127 & 0.13 & 0.21 &
M028 & 0.1527 & 0.0222 & 0.8694 & 0.932 & 0.8068 & 0.6189 & 0.11 & 0.18 \\ 
M010 & 0.1403 & 0.0218 & 0.8565 & 0.990 & 0.7556 & 0.6695 & 0.12 & 0.19 &
M029 & 0.1256 & 0.0228 & 1.0435 & 0.913 & 0.7087 & 0.7067 & 0.13 & 0.21 \\ 
M011 & 0.1437 & 0.0234 & 0.8823 & 1.126 & 0.7276 & 0.7177 & 0.13 & 0.21 &
M030 & 0.1234 & 0.0230 & 0.8758 & 0.777 & 0.6739 & 0.6626 & 0.12 & 0.19 \\ 
M012 & 0.1223 & 0.0225 & 1.0048 & 0.971 & 0.6271 & 0.7396 & 0.15 & 0.24 &
M031 & 0.1550 & 0.0219 & 0.9919 & 1.068 & 0.7041 & 0.6394 & 0.13 & 0.23 \\ 
M013 & 0.1482 & 0.0221 & 0.9597 & 0.855 & 0.6508 & 0.6107 & 0.14 & 0.23 &
M032 & 0.1200 & 0.0229 & 0.9661 & 1.048 & 0.7556 & 0.7901 & 0.13 & 0.19 \\
M014 & 0.1471 & 0.0233 & 1.0306 & 1.010 & 0.7075 & 0.6688 & 0.14 & 0.23 &
M033 & 0.1399 & 0.0225 & 1.0407 & 1.147 & 0.8645 & 0.7286 & 0.12 & 0.19 \\ 
M015 & 0.1415 & 0.0230 & 1.0177 & 1.281 & 0.7692 & 0.7737 & 0.14 & 0.22 &
M034 & 0.1497 & 0.0227 & 0.9239 & 1.000 & 0.8734 & 0.6510 & 0.11 & 0.18 \\  
M016 & 0.1245 & 0.0218 & 0.9403 & 1.145 & 0.7437 & 0.7929 & 0.13 & 0.20 &
M035 & 0.1485 & 0.0221 & 0.9604 & 0.853 & 0.8822 & 0.6100 & 0.10 & 0.17 \\ 
M017 & 0.1426 & 0.0215 & 0.9274 & 0.893 & 0.6865 & 0.6305 & 0.13 & 0.21 &
M036 & 0.1216 & 0.0233 & 0.9387 & 0.706 & 0.8911 & 0.6421 & 0.09 & 0.15 \\
M018 & 0.1313 & 0.0216 & 0.8887 & 1.029 & 0.6440 & 0.7136 & 0.14 & 0.23 &
M037 & 0.1495 & 0.0228 & 1.0233 & 1.294 & 0.9000 & 0.7313 & 0.12 & 0.19
\end{tabular}
\end{center}
\label{tab:basic}
\end{table*}

This paper is the third in a series aimed at addressing this question
in the context of the nonlinear matter power spectrum on Mpc scales.
Current observations in weak lensing are quickly becoming theory
limited due to the lack of precise theoretical estimates of this quantity
for a wide range of cosmological models. This is a pressing problem.
It is also relatively simple, allowing us to work through -- in a
concrete setting -- the many steps which will be routinely required in
the future. In some ways the problem is one of the simplest currently
confronting theorists, but as we shall discuss below even it has
demanded collaboration with other communities, the development of a
significant infrastructure, new modes of working, and large amounts of
manpower and computational capacity.

In Paper~I of this series (Heitmann et al. 2009a) we demonstrated that
it is possible to obtain nonlinear matter power spectra with percent
level accuracy out to $k\simeq 1\,h\,{\rm Mpc}^{-1}$ and derived a set
of requirements for such simulations. Paper~II \citep{CoyoteII}
described the construction of an emulation scheme to predict the
nonlinear matter power spectrum and the underlying cosmological
models, building on the ``Cosmic Calibration Framework''
\citep{HHHN,HHHNW,SKHHHN}. Here, in Paper~III, we present
results from the complete simulation suite based on the cosmologies
presented in the second paper and publicly release a precision power
spectrum
emulator~\footnote{http://www.lanl.gov/projects/cosmology/CosmicEmu}. The
simulation suite is called the ``Coyote Universe'' after the cluster
it has been carried out on. We will extend our work to include other
measurements, such as the mass function or higher order statistics, in
future publications.

The outline of this paper is as follows.  In \S\ref{sec:sims} we
describe the simulations we have run in support of this program, and
the codes with which they were run.  \S\ref{sec:power} describes how
we put together the different estimations of the power spectra from
our multi-scale runs while \S\ref{sec:emulator} presents the details
of the emulator and tests.  We describe some lessons learned in
\S\ref{sec:lessons} before concluding in \S\ref{sec:conclusions}.

\section{The Simulation Suite} \label{sec:sims}

\subsection{The Simulations}

The Coyote Universe simulation suite encompasses nearly 1,000
simulations of varying force and mass resolution.  The simulation
volume is the same in all cases, a periodic cube of side length
$1300\,$Mpc.  We consider 37+1 cosmological models, listed in Table
\ref{tab:basic}, which we select with two aspects in mind: our
statistical framework and current constraints from a variety of
cosmological measurements (see Paper II for further discussions and
see below for a short summary).

The 37 models, labeled 1-37, are used to construct the emulator while
model M000 is used as an independent check on the power spectrum
accuracy in the parameter regime of most interest. (Other tests are
described in \S\ref{sec:emulator} and in Paper~II.)

For each cosmology we run 20 realizations, 16 lower resolution
simulations covering the low-$k$ regime (which we refer to as the
``L-series''), 4 medium resolution runs to provide good statistics in
the quasi-linear to mildly nonlinear regime (the ``H-series''), and
one high resolution run to extend to $k\simeq 1\,h\,{\rm Mpc}^{-1}$
(the ``G-series''). The high-resolution run uses the same realization
as one of the medium resolution runs. The low- and medium-resolution
runs are performed with a particle-mesh (PM) code. The code evolves $512^3$
particles in the low-resolution runs and $1024^3$ particles in the
medium-resolution runs, in each case with a force mesh twice as large
in each dimension as the particle load. Densities and forces are
computed using Cloud-in-Cell (CIC) interpolation and a Fast Fourier
Transform (FFT)-based Poisson solver. The potential is determined from
the density using a $1/k^2$ Green's function and the force is computed
by $4^{\rm th}$ order differencing. Particles are advanced with second
order (global) symplectic time-stepping with $\Delta\ln a=2\%$. In
order to resolve the high-$k$ part of the power spectrum, we evolve
one of the $1024^3$ particle initial distributions with the
Tree-PM code {\sc GADGET-2\/} \citep{Spr05}. We use a PM
grid twice as large, in each dimension, as the number of particles,
and a (Gaussian) smoothing of $1.5$ grid cells. The force matching is
set to $6$ times the smoothing scale, the tree opening criterion to
$0.5\%$ and the softening length to $50\,$kpc. The starting redshift
for each simulation is $z=211$. Further details regarding the
simulations and choices of simulation parameters can be found in Paper
I.

All the simulations are carried out on {\it Coyote}, a large HPC Linux
cluster consisting of 2580 AMD Opterons running at $2.6\,$GHz.  The
low-resolution PM runs are carried out on 64 processors, the
medium-resolution PM runs and {\sc GADGET-2\/} runs on 256.  For the
billion particle runs we store particle and halo information at 11
different redshifts between $z=4$ and $z=0$.  This leads to a final
simulation database of size roughly $60\,$TB.

\subsection{The Cosmological Models}

The selection of the cosmological model suite depends on two
considerations: the statistical framework we use to construct the
emulator and current parameter constraints from the CMB as set by
WMAP-5~\citep{WMAP5}. We do not insist on a formal methodology to make
the model selection, but instead apply some practical and conservative
arguments to justify our decisions. For an in-depth discussion we
refer the reader to Paper II. In this paper we briefly summarize some
of the considerations since these will define the region for which the
emulator will be valid.

Our aim is to find a distribution of the parameter settings -- the
simulation design -- which provides optimal coverage of the parameter
space, using only a limited number of sampling points. Simulation
designs well-suited for this task are Latin-Hypercube (LH)-based
designs, a type of stratified sampling scheme. Latin hypercube
sampling generalizes the Latin square for two variables, where only
one sampling point can exist in each row and each column. A Latin
hypercube sample -- in arbitrary dimensions -- consists of points
stratified in each (axis-oriented) projection.

Very often LH designs are combined with other design strategies such
as orthogonal array (OA)-based designs or are optimized in other ways,
e.g., by symmetrizing them (more details below). By intelligently melding
design strategies, different attributes of the individual sampling
strategies can be combined to lead to improved designs, and shortcomings of 
specific designs can be eliminated. As a last step, optimization
schemes are often applied to spread out the points evenly in a
projected space. One such optimization scheme is based on minimizing
the maximal distance between points in the parameter space, which will
lead to more even coverage. Two design strategies well suited to
cosmological applications in which the number of parameters is much
less than the number of simulations that can be performed are
optimal OA-LH design strategies and optimal symmetric LH design
strategies. For this project, we have generated forty different
designs following different strategies and chose the best design from
these (where ``best'' refers to best coverage in parameter space with
respect to specific distance criteria explained in Paper II). 

In order to restrict the number of necessary simulation runs to an as
small number as possible, it is helpful to keep the number of
cosmological parameters and their prior ranges small. Current
observations of the CMB and the large scale structure are consistent
with a $\Lambda$CDM model with constant dark energy equation of state,
$w$. We therefore concentrate on the following five cosmological
parameters: $\omega_m\equiv\Omega_m h^2$, $\omega_b\equiv\Omega_b
h^2$, $n_s$, $w$, and $\sigma_8$ where $\Omega_m$ contains the
contributions from the dark matter and the baryons.  We restrict
ourselves to power-law models (no running of the spectral index) and
to spatially flat models without massive neutrinos.  For each
cosmology, $h$ is determined by the angular scale of the acoustic
peaks in the CMB (Paper II) which is known to very high accuracy
\citep[0.3\%;][]{WMAP5}. From WMAP 5-year data, in combination with
BAO, we have\footnote{See http://lambda.gsfc.nasa.gov} 
\begin{equation}
\begin{array}{c}
  \omega_m = 0.1347\pm 0.0040\quad (3\%), \\
  \omega_b = 0.0227\pm 0.0006\quad (3\%), \\
  n_s      = 0.9610\pm 0.0140\quad (2\%).
\end{array}
\end{equation}
Current data restrict a constant equation of state for the dark energy
to $w=-1$ with roughly 10\% accuracy and recent determinations put the
normalization in the range $0.7 < \sigma_8 < 0.9$ with still rather large
uncertainties.

Considering all these constraints and their uncertainties, we choose
our sample space boundaries for the 37+1 models to lie within the
range(s)
\begin{equation}
\begin{array}{c}
  0.120 < \omega_m < 0.155, \\
  0.0215 < \omega_b < 0.0235, \\
  0.85  < n_s      < 1.05, \\
  -1.30 < w        <-0.70, \\
  0.61  < \sigma_8 < 0.9,
\end{array}
\label{priors}
\end{equation}
over which the emulator is designed to produce reliable results.  We
verified in Paper II that 37 models spanning these parameter ranges
are indeed enough to generate an emulator at the 1\% accuracy. We
emphasize that our emulator is valid for the {\em complete\/}
parameter space defined by these priors and not restricted to some
values in a band around the best-fit cosmology (for current
observational data).  Obviously, the emulator quality will be slightly
worse on the edges of the hypercube but the emulator is always
accurate within $\sim$1\% anywhere within the priors.

\subsection{Power Spectra}

In Paper I we describe in detail how we obtain the matter
power spectrum from a snapshot of the simulation. We briefly summarize
the salient points here.  We compute the dimensionless power spectrum,
\begin{equation}
  \Delta^2(k) \equiv \frac{k^3P(k)}{2\pi^2},
\label{eqn:delta}
\end{equation}
which is the contribution to the variance of the density perturbations
per $\ln k$.  We obtain $\Delta^2$ by binning the particles onto a
2048$^3$ grid using CIC assignment, applying an FFT, correcting for the
charge-assignment window function, and averaging the result in fine
bins in $|\mathbf{k}|$ spaced linearly with width $\Delta k\simeq
0.001\,{\rm Mpc}^{-1}$.  As discussed in Paper I, we do not
correct for particle discreteness as our particle loading is high
enough to make such corrections unnecessary and there are some
indications that a simple Poisson shot-noise form is not correct.

\section{Power Spectrum Determination} \label{sec:power}

When creating a nonlinear power spectrum emulator, we prefer
that the underlying training data set be smooth.  We describe here how
we construct a smooth power spectrum from the 20 realizations of each
cosmology and from cosmological perturbation theory.

In each simulation the modes in the initial density field are a single realization of a Gaussian
random field and this introduces large run-to-run scatter. This scatter is
reduced at higher $k$ by the relatively large number of modes which
are averaged. However, at low $k$ the estimates of the power spectrum
exhibit significant scatter which is expensive to reduce by brute
force, i.e. running a very large number of realizations in large volumes. In addition, the approach to linear theory at low $k$ can be
quite slow for many currently popular models around $\Lambda$CDM (if
$1\%$ accuracy is the desired goal) so simply replacing the N-body
results with the input linear model can be relatively inaccurate. This
is however an area where perturbation theory can be of help, since the
real-space mass power spectrum is computable in perturbation theory
and there is a small but non-negligible range of scales where
perturbation theory improves upon linear theory. For a recent overview
of the performance of different perturbation theory approaches, see
\cite{carlson09}.

\subsection{Perturbation theory}

\begin{figure}[t] \begin{center}
    \resizebox{2.5in}{!}{\includegraphics{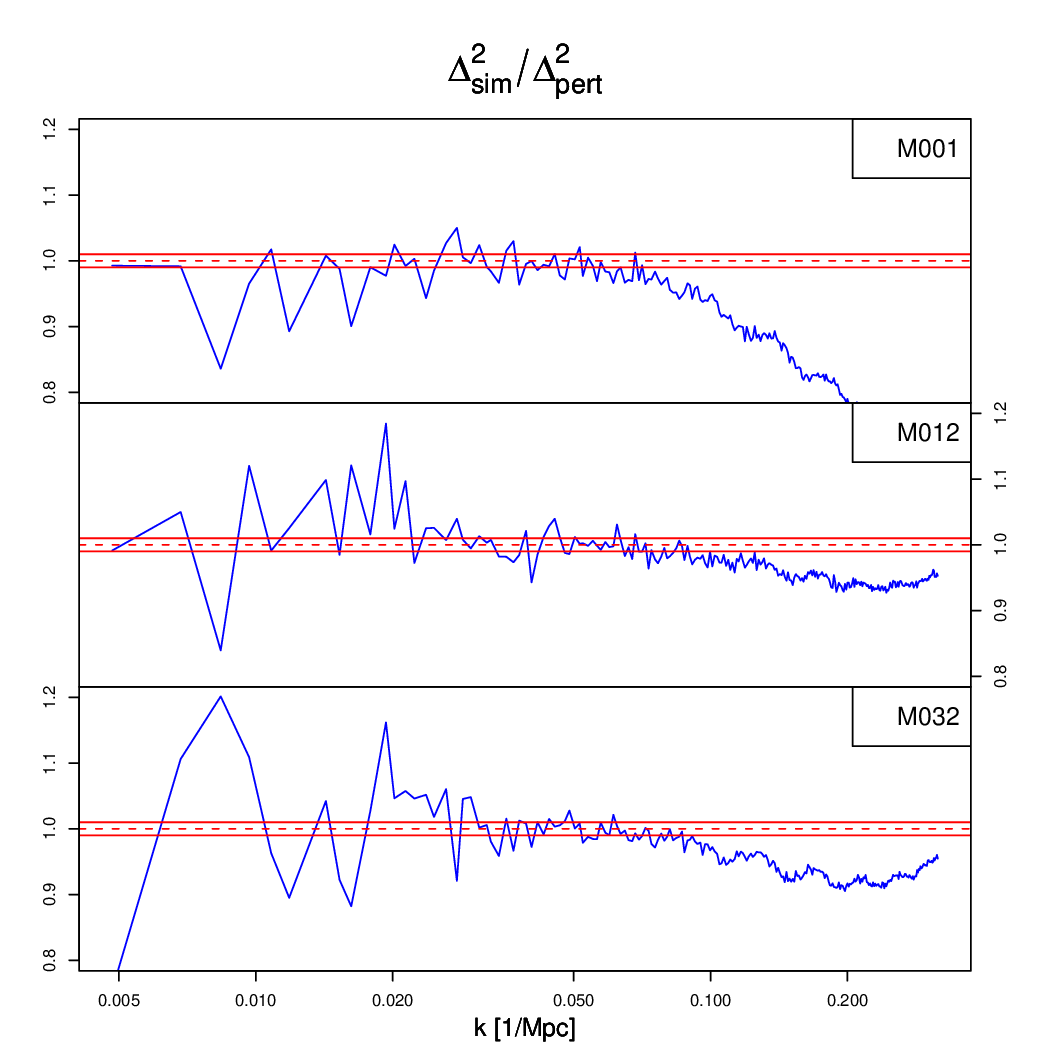}} \end{center}
  \caption{Comparison of perturbation theory with the nonlinear matter
    power spectrum from simulations at $z=0$. Out of our 38 models we
    show three random examples, the results for the remaining models being
    very similar. For each model we determine a simulated power
    spectrum by matching the sixteen lower resolution L runs to the
    four medium resolution H runs at $k=0.25$~Mpc$^{-1}$. Therefore we
    average over 20 (1.3Gpc)$^3$ simulations at low $k$-values and
    four simulations beyond $k=0.25$~Mpc$^{-1}$. We then take the
    ratio of the simulation results with respect to perturbation
    theory. The agreement is very good out to at least $k\simeq
    0.03~$Mpc$^{-1}$. We therefore conclude that -- up to these wavenumbers 
    -- perturbation theory results are robust for the cosmologies
    considered in this paper.}
\label{fig:compare}
\end{figure}

While there has been a resurgence of interest in perturbative methods
in recent years, we stick to the simplest and oldest method:
``standard perturbation theory''
\citep{Pee80,Jus81,Vis83,Gor86,Mak92,JaiBer94}.  We consider only the
first correction to linear theory, which in standard perturbation
theory can be written in terms of a simple integral over the linear
theory power spectrum (we use the specific form given in
\citealt{MWP99}).  When compared to our simulations, we find that
standard perturbation theory is accurate at the percent level for
$k<0.5\,k_{\rm nl}$ where $k_{\rm nl}$ can be defined as
\citep{mats08}:
\begin{equation}
  k_{\rm nl}^{-2} = \frac{1}{3}\int \frac{dk}{k}\ \frac{\Delta^2(k)}{k^2}.
\end{equation}
The values for $k_{\rm nl}$ at $z=0$ and $z=1$ are listed for all
models in Table~\ref{tab:basic}. For example, for model M000, $k_{\rm
  nl}\simeq 0.1\,{\rm Mpc}^{-1}$ at $z=0$.  Similar results have been
reported in \citet{mats08} and \citet{carlson09}.

Almost any scheme that switches smoothly from standard perturbation
theory to our N-body results around $0.5\,k_{\rm nl}$ produces results
that agree at the percent level. At $z=0$, this would lead to a
matching point between $k=0.045$~Mpc$^{-1}$ (M036) and
$k=0.075$Mpc$^{-1}$ (M019). For simplicity we keep the matching point
the same for all cosmologies. Since we have good statistics from our
simulations at $k\simeq 0.03$~Mpc$^{-1}$ already, we choose this $k$
value as a very conservative matching point for all models. To verify
this point, we compare our simulation results to perturbation theory
for the different cosmologies. The simplest approach for this
comparison is ``brute force'': simply take the ratio of the simulation
result to the perturbation theory prediction. The major obstacle here
is the run-to-run scatter in the simulations. In order to overcome
this problem, one can follow two routes: either incorporate
fluctuations from the realization into the perturbation theory
prediction, as done in \cite{taka08}, or average over a large number
of large volume simulations. We follow the second approach to avoid
any ambiguities and systematic errors resulting from finite box size
effects.

An example of our approach is shown in Figure~\ref{fig:compare} for a
random subset of 3 models from the total of 37 used to build the
emulator. Although at the lower end of the $k$ range, the large but
finite sampling volume becomes clearly evident, the matching to
perturbation theory at $k\simeq 0.03~$Mpc$^{-1}$ works extremely well.
A similar result can also be found in \cite{carlson09}.

\subsection{The Estimation Procedure}
\label{smooth}

Next, we show how we can combine perturbation theory and results from
$N$-body simulations to generate smooth power spectra which will be
the foundation for building our emulator. Two problems have to be
solved for this: (1) We have to eliminate the scatter in the $N$-body
results for the nonlinear power spectrum without erasing subtle
features and match results from different resolution simulations. (2)
We have to match very accurately between perturbation theory and
simulation results. In the following, we discuss an approach
based on process convolution to solve these problems.

\subsubsection{Power Spectrum Estimation using Process Convolution}
\begin{figure}[t]
\begin{center}
\resizebox{2.5in}{!}{\includegraphics{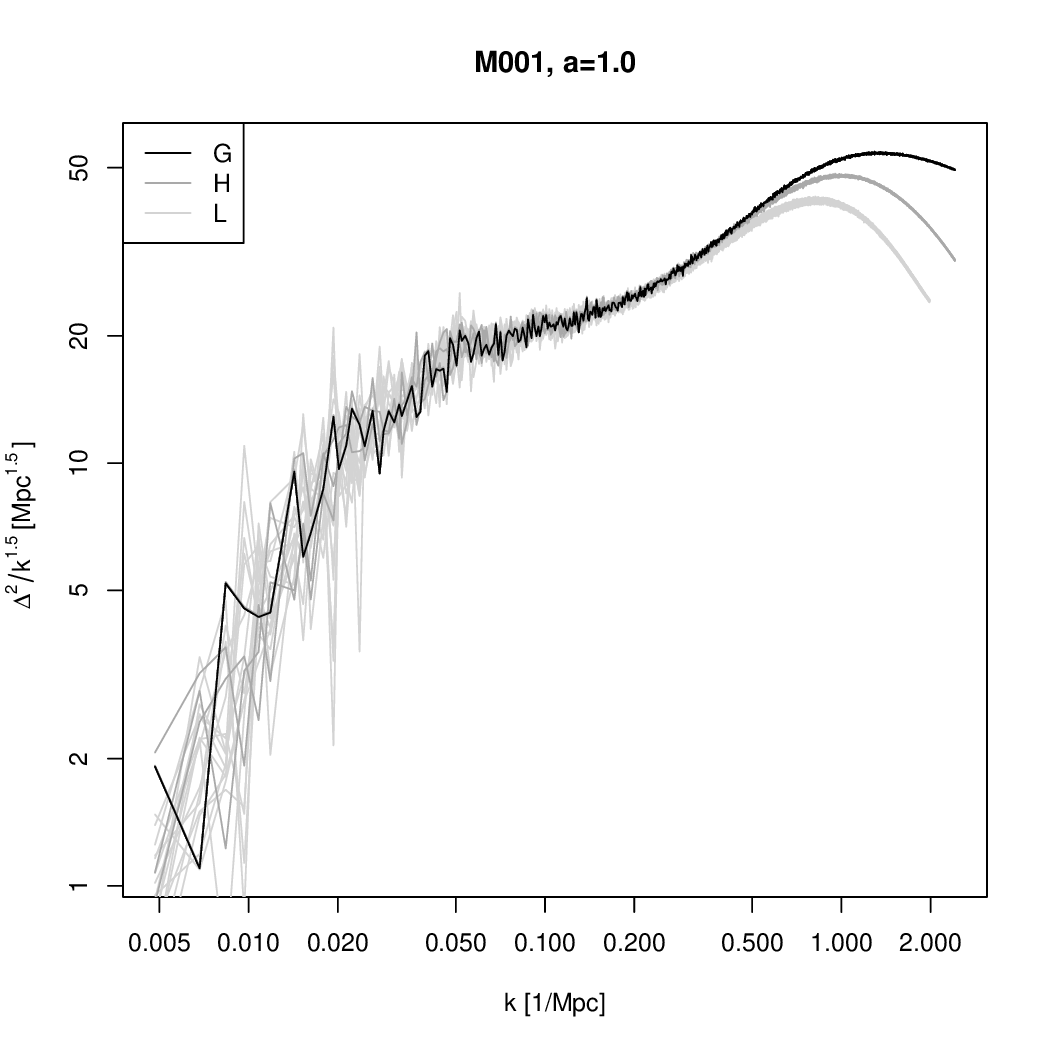}}
\end{center}
\caption{Power spectra from the $N$-body simulations for cosmology
M001 on the scales that are used for smoothing.  The insufficient
force resolution of the PM runs (H and L) is apparent at large $k$. To
generate a smooth power spectrum over the whole $k$ range, we match
perturbation theory and the L-runs at $k=0.03$~Mpc$^{-1}$, the L- and
H-series at $k=0.25$~Mpc$^{-1}$, and use the G-series for $k\ge
0.32$~Mpc$^{-1}$. Note the non-standard representation of the power spectrum
via $\Delta^2(k)/k^{1.5}$.}
\label{fig:M001}
\end{figure}

In this section, we discuss the procedure for estimating the smooth
power spectrum for each cosmology based on the simulation results
which possess inherent scatter. Figure \ref{fig:M001} shows the power
spectrum from the simulations for model M001 at $z=0$. The data have
three notable features that are important for the modeling procedure.
(1) The non-standard representation for the power spectrum
\begin{equation}
{\cal P}(k)=\Delta^2(k) /k^{1.5}
\end{equation}
is chosen to accentuate the baryon acoustic oscillations. We will
account for this feature when we choose our function class for the
smooth power spectra (discussed later). (2) The three series of
simulations; G, H, and L with 1, 4, and 16 realizations respectively.
Because the H and L series do not have enough force resolution to
resolve the nonlinear regime at high $k$, we need to restrict the use
of these runs to small and intermediate $k$ ranges. (3) The simulation
variance at any given $k$ is known.

We treat each simulated (``noisy'') realization of the (large volume or
averaged) spectrum as a draw from a multivariate
Gaussian distribution whose mean is given by an unknown smooth
spectrum.  Thus, for a given cosmology $c$, a given series $s \in
\{\mbox{G}, \mbox{H}, \mbox{L}\}$, and a given replicate $i = 1,
\cdots, N_s$ (where $N_s$ is the number of simulations for the
series), we have a multivariate Gaussian density for the simulated
spectrum $P^{c}_{s,i}$,
\begin{eqnarray}
f(P^{c}_{s,i}) & \propto & \left| A_s \Omega A_s^{T} \right|^{1/2} \\
& \times & \exp \left\{ -\frac{1}{2} \left(P^{c}_{s,i} - A_s
\mathcal{P}^{c} \right)^{T} A_s \Omega A_s^{T}
\left(P^{c}_{s,i} - A_s \mathcal{P}^{c} \right) \right\}. \nonumber
\end{eqnarray}
Here, $\mathcal{P}^c$ is the smooth power spectrum; $A_s$ is a projection
matrix of zeros and ones that is used to remove the high-$k$ values
for which the H and L series are not used (thus, $A_G$ is an
identity matrix); and $\Omega$ is a diagonal matrix of the known
precisions (inverse variances).

\begin{figure}[t]
\begin{center}
\includegraphics{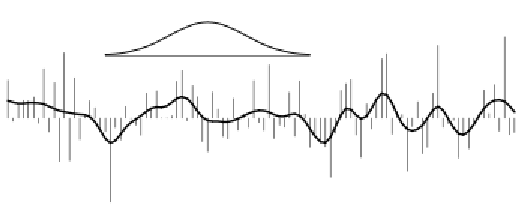}
\end{center}
\caption{The process convolution model treats a smooth function as arising from  
a weighted average of a simple stochastic process.  In this figure,
each point on the smooth function is a weighted average of the
Gaussian impulses shown as the vertical bars.  In these case, the
weight function is a Gaussian kernel.  Typically, the smooth function
is observed (possibly with noise) and the challenge is to estimate the
impulses and perhaps some aspect of the smoothing kernel.}
\label{fig:proc-conv}
\end{figure}

The model is completed by specifying a class of functions for the
smooth power spectrum $\mathcal{P}^c$.  We choose a flexible class of
smooth functions called a process convolution \citep{Higdon2002}.
These functions are best described constructively.  A process
convolution builds a smooth function as a moving average of a simple
stochastic process like independent and identically-distributed ({\em
i.i.d.})~Gaussian variates or Brownian motion.  The moving average
uses a smoothing kernel whose width is allowed to vary over the domain
to account for nonstationarity (smoother in some regions, more wiggly
in others).  Figure \ref{fig:proc-conv} shows a simple example of a
process convolution built on white noise smoothed with a Gaussian
kernel.

We build the process convolution for $\mathcal{P}^c$ on Brownian
motion $u^c$, with marginal variance $\tau^2_u$, realized on a sparse
grid (relative to the power spectrum) of evenly spaced points, $x$
(the number of points is not important so long as it is large enough; we
use 100), 
\begin{equation} \label{eq:uprior}
f(u^c) \propto \left| \frac{1}{\tau^2_u} W \right|^{1/2} \exp
\left\{ -\frac{u^{c \prime} W u^c}{2 \tau^2_u} \right\},
\end{equation}
where $W$ is the Brownian precision matrix with diagonal equal to
$[1~2~\cdots~2~1]$ and $-1$ on the first off-diagonals (this matrix
cannot actually be inverted, but never has to be in the estimation).
The Brownian motion is transformed into $\mathcal{P}^c$ by the
smoothing matrix $K^{\sigma}$,
\begin{equation} \label{procsmooth}
\mathcal{P}^c = K^{\sigma} u^c.
\end{equation}
The smoothing matrix $K^{\sigma}$ is built using Gaussian kernels
whose width varies smoothly across the domain.  Thus, we have
\begin{equation}
K^{\sigma}_{i,j} = \frac{1}{\sqrt{2 \pi \sigma^{2}_{i}}} \exp \left\{
-\frac{ \left( \log_{10}(k_i) - x_j \right)^2}{2 \sigma_i^2} \right\},
\end{equation}
where $k_i$ is the $i$-th value of $k$ for which the power spectrum is
computed.

In the description of $K^{\sigma}$, $\sigma$ is indexed to indicate
that it changes over the domain.  Intuitively, we want $\sigma$ to be
small in the middle of the domain in order to capture the oscillations,
but large elsewhere to smooth away the noise.  In order to estimate
this varying bandwidth parameter, we build a second process
convolution model.  This model is built on {\em i.i.d.} Gaussian
variates $v$, with mean zero and variance $\tau^2_v$, observed on an
even sparser grid of evenly spaced points,
$t$ (length $M_v$; we use 10, but any large enough number will suffice),
\begin{equation} \label{eq:vprior}
f(v) \propto \left( \frac{1}{\sqrt{ \tau_{v}^{2} } } \right)^{M_v}
\exp \left\{ -\frac{v^{\prime} v}{2 \tau^2_v} \right\}.
\end{equation}
The process $v$ is transformed into $\sigma$ by the smoothing matrix
$K^{\delta}$, 
\begin{equation}
\sigma = K^{\delta} v.
\end{equation}

\begin{figure*}
\begin{center}
\resizebox{2.2in}{!}{\includegraphics{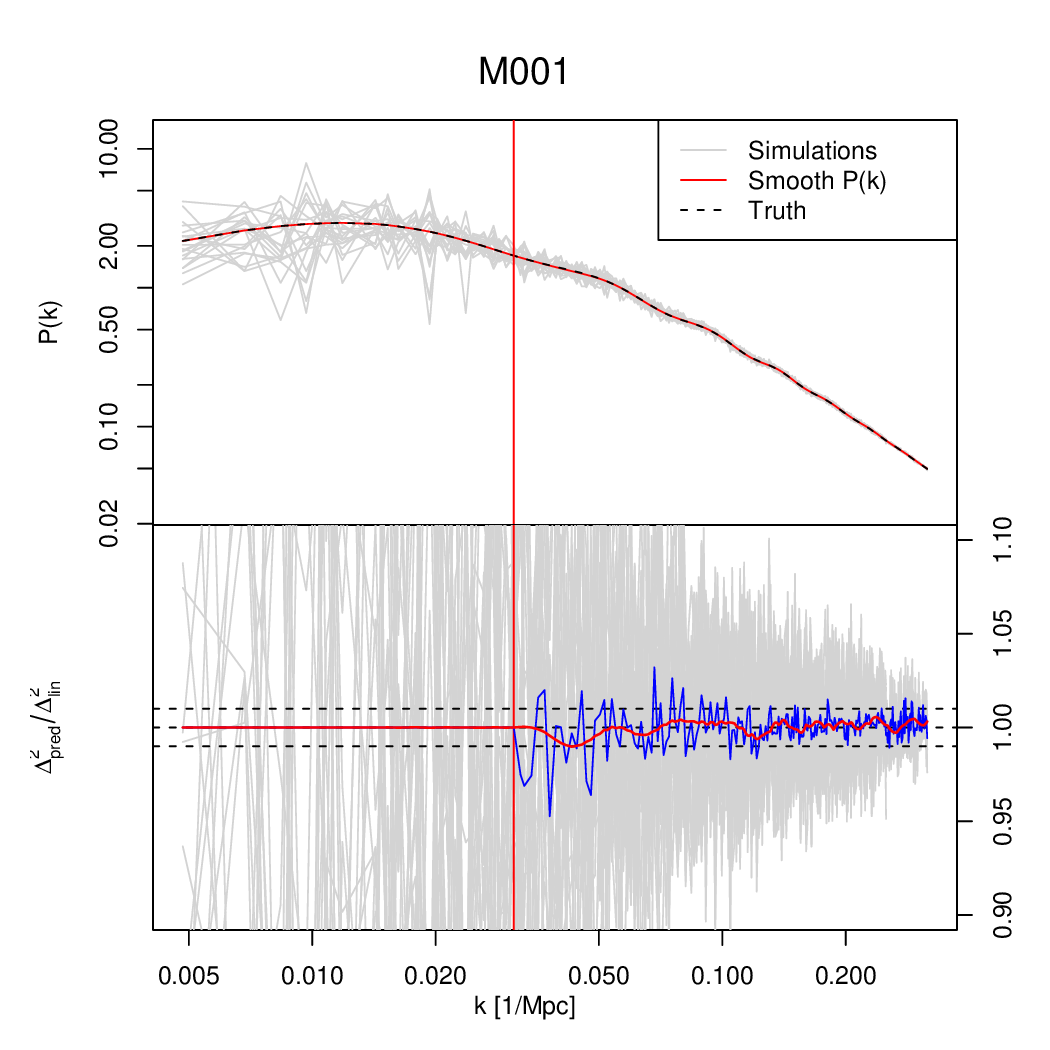}}
\resizebox{2.2in}{!}{\includegraphics{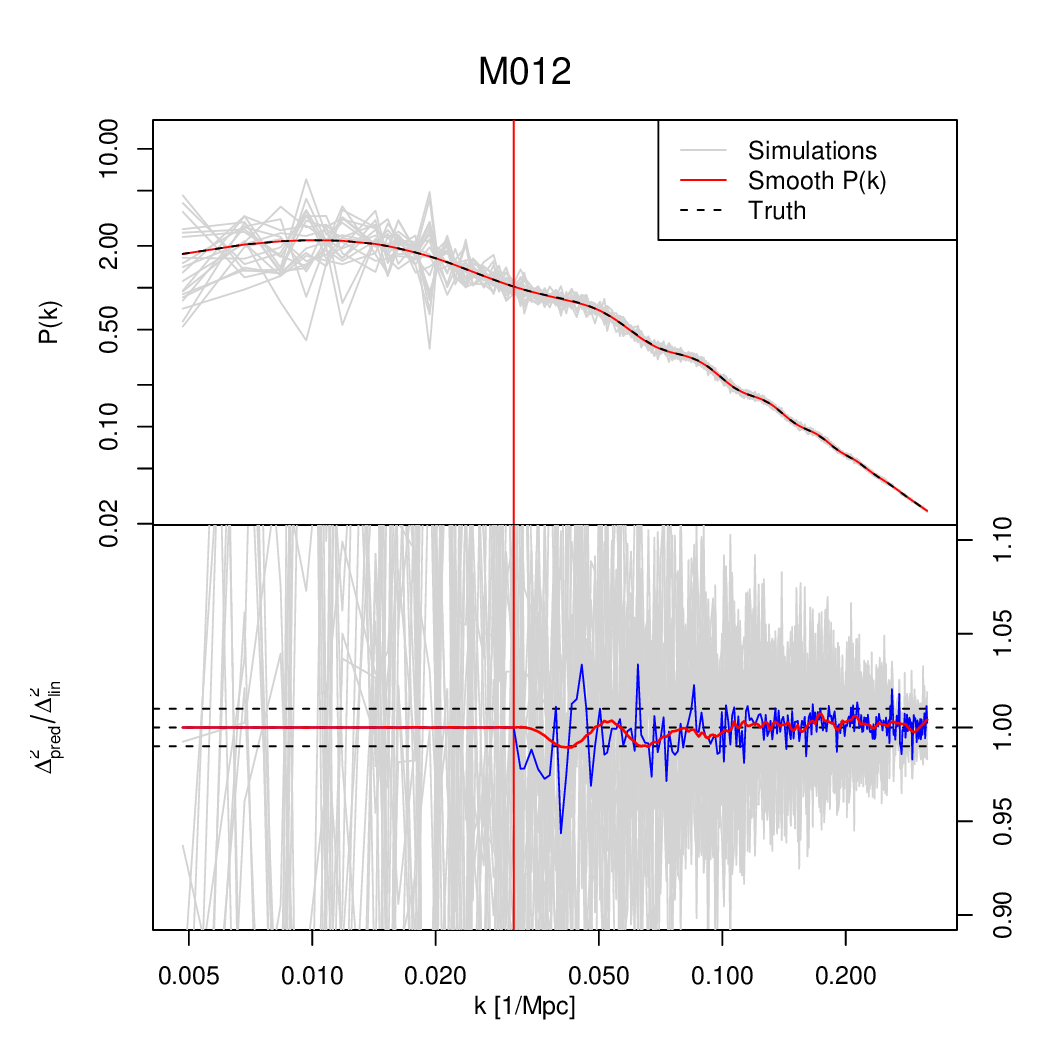}}
\resizebox{2.2in}{!}{\includegraphics{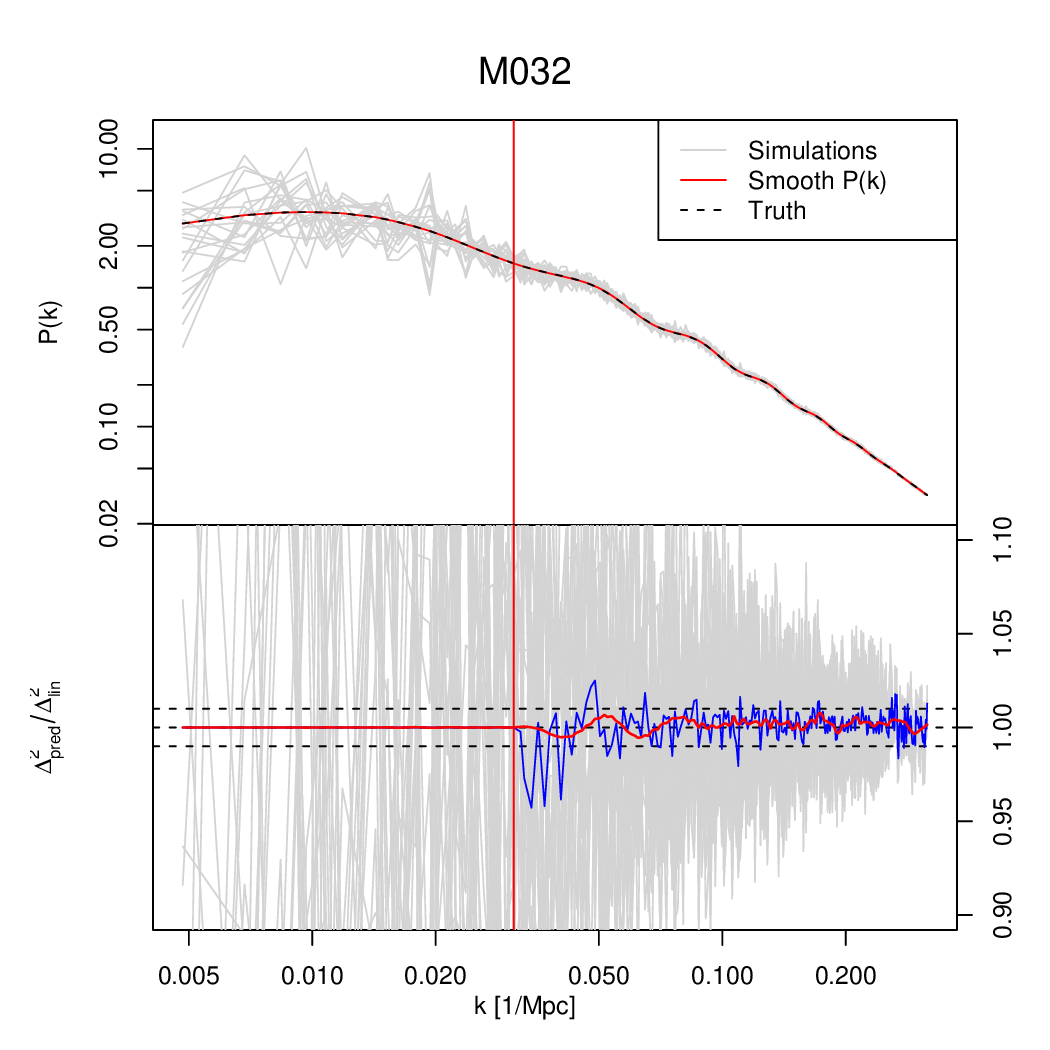}}
\end{center}
\caption{Predictions for the linear power spectrum from 20
  realizations of the initial power spectrum for the same three models
  shown in Figure~\ref{fig:compare}. The upper panels for each model
  show in gray the realizations, in red the smooth power spectrum and in black
  (dashed) the linear theory power spectrum that underlies the
  simulations. The lower panels show the realizations
  (again in gray) divided by the linear theory answer and in red the
  smooth power spectrum divided by linear theory. The blue line shows the
  empirical mean of the realizations divided by linear theory. The
  vertical red lines in each panel indicate the matching point of
  linear theory and simulation result, the horizontal dashed lines in
  the lower panels show the 1\% error. In all models the discrepancy
  is at the 1\% level.}
\label{fig:lin_pred}
\end{figure*}

This matrix is also built using Gaussian smoothing kernels, but
with a constant bandwidth, $\delta$.  Thus, we have 
\begin{equation}
K^{\delta}_{i,j} = \frac{1}{\sqrt{2 \pi \delta^{2}}} \exp \left\{
  -\frac{\left(x_i - t_j  \right)^2}{2 \delta^2}  \right\}.
\end{equation}

Combining all of this, we get a distribution for the simulated power
spectra for a given cosmology, 
\begin{eqnarray} \label{eq:likelihood}
f(P^{c}_{s,i}) & \propto & \left| A_s \Omega A_s^{\prime}
\right|^{1/2} \\ & \times & \exp \left\{ -\frac{1}{2}
\left(P^{c}_{s,i} - A_s K^{\sigma} u^c \right)^{\prime} A_s \Omega
A_s^{\prime} \left(P^{c}_{s,i} - A_s K^{\sigma} u^c \right)
\right\}, \nonumber
\end{eqnarray}
where $s \in \{\mbox{G}, \mbox{H}, \mbox{L}\}$ and $i = 1, \cdots,
N_s$. Note that $K^{\sigma}$ is a function of the parameters $v$ and
$\delta$.  We choose noninformative priors for $\tau^{2}_{u}$,
$\tau^2_v$, and $\delta$ so as to impart little or no information
about their values,
\begin{eqnarray} \label{eq:priors}
\pi(\tau^2_u) & \propto & \left( \frac{1}{\tau^2_u} \right)^2 \exp
\left(-\frac{.001}{\tau^2_u} \right),  \nonumber \\
\pi(\tau^2_v) & \propto & \left( \frac{1}{\tau^2_v} \right)^2 \exp
\left(-\frac{.001}{\tau^2_v} \right),  \\
\pi(\delta) & \propto & I\{ 0 \leq \delta \leq 10 \}, \nonumber
\end{eqnarray}
which are inverse gamma, inverse gamma, and uniform, respectively.

Equations (\ref{eq:uprior}) (for all $c$), (\ref{eq:vprior}), and
(\ref{eq:priors}) are multiplied together with
Eqn.~(\ref{eq:likelihood}) (for all $c$) to produce a posterior
distribution for the unknown parameters and stochastic processes.
Obtaining an estimate for the smooth power spectra requires an
estimate for each of the parameters, the process $v$, and each of the
processes $u^c$ for all $c$. Markov Chain Monte Carlo (MCMC) via the
Metroplis-Hastings algorithm \citep{ChibGreenberg1995} produces a
sample from the posterior distribution by drawing each parameter
individually. The stochastic processes $u^c$ can be integrated out of
the distribution, so the MCMC produces samples of the parameters as
well as $v$. We use the posterior mean of the parameters to obtain a
conditional mean for each of the $u^c$ which is then transformed into
each of the $\mathcal{P}^c$.

The perturbation results are included by setting the low $k$ values
for each of the simulated spectra in a given cosmology to the
perturbation results and setting the precisions for these values to be
very large.  This is done prior to the estimation described here.  The
replication of these values in every simulation, combined with the
large precisions, nearly forces the estimated result through these
points.  Further, the transition from the N-body results to the
perturbation results will be relatively smooth as long as the two line
up fairly well.

\subsubsection{Tests with the Linear Power Spectrum}

\begin{figure*}[t]
\begin{center}
\resizebox{2.34in}{!}{\includegraphics{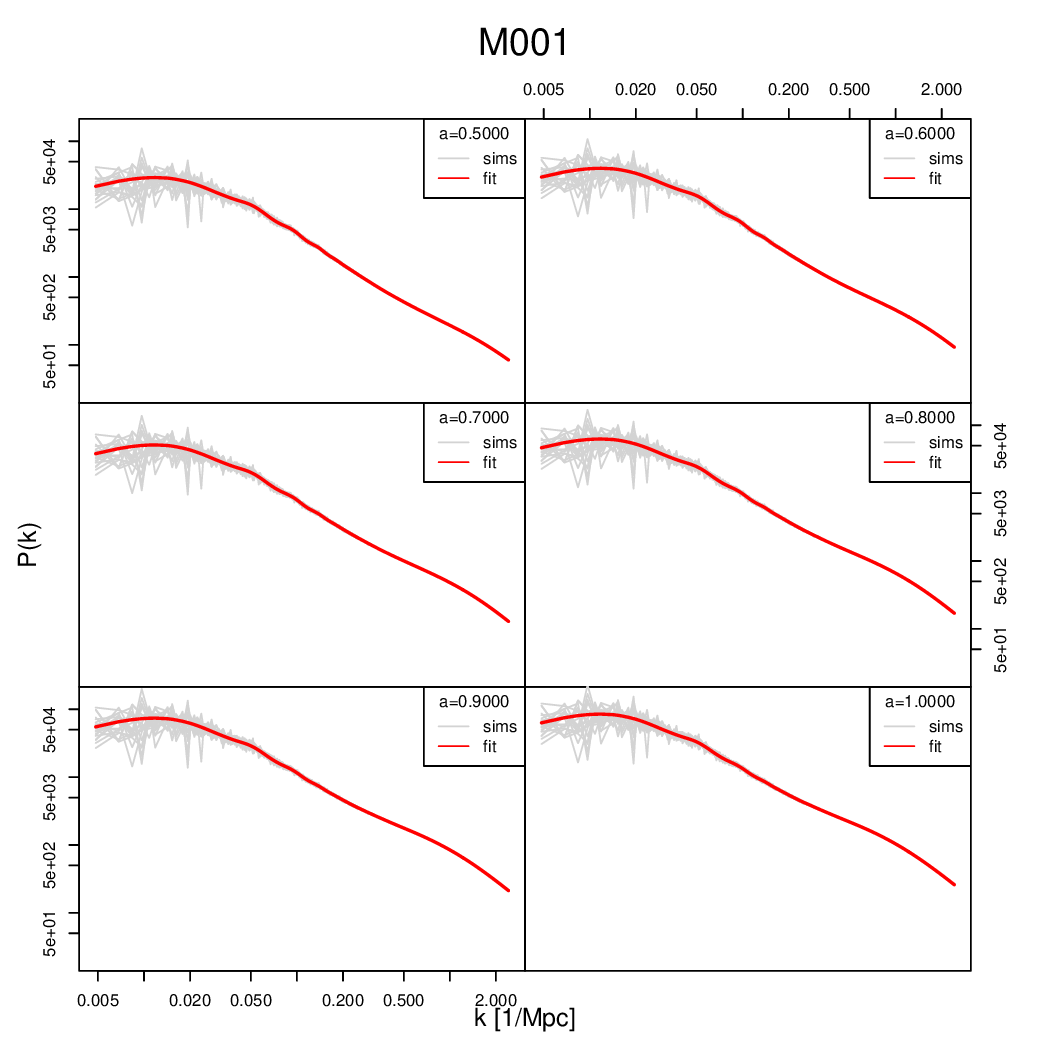}}
\resizebox{2.34in}{!}{\includegraphics{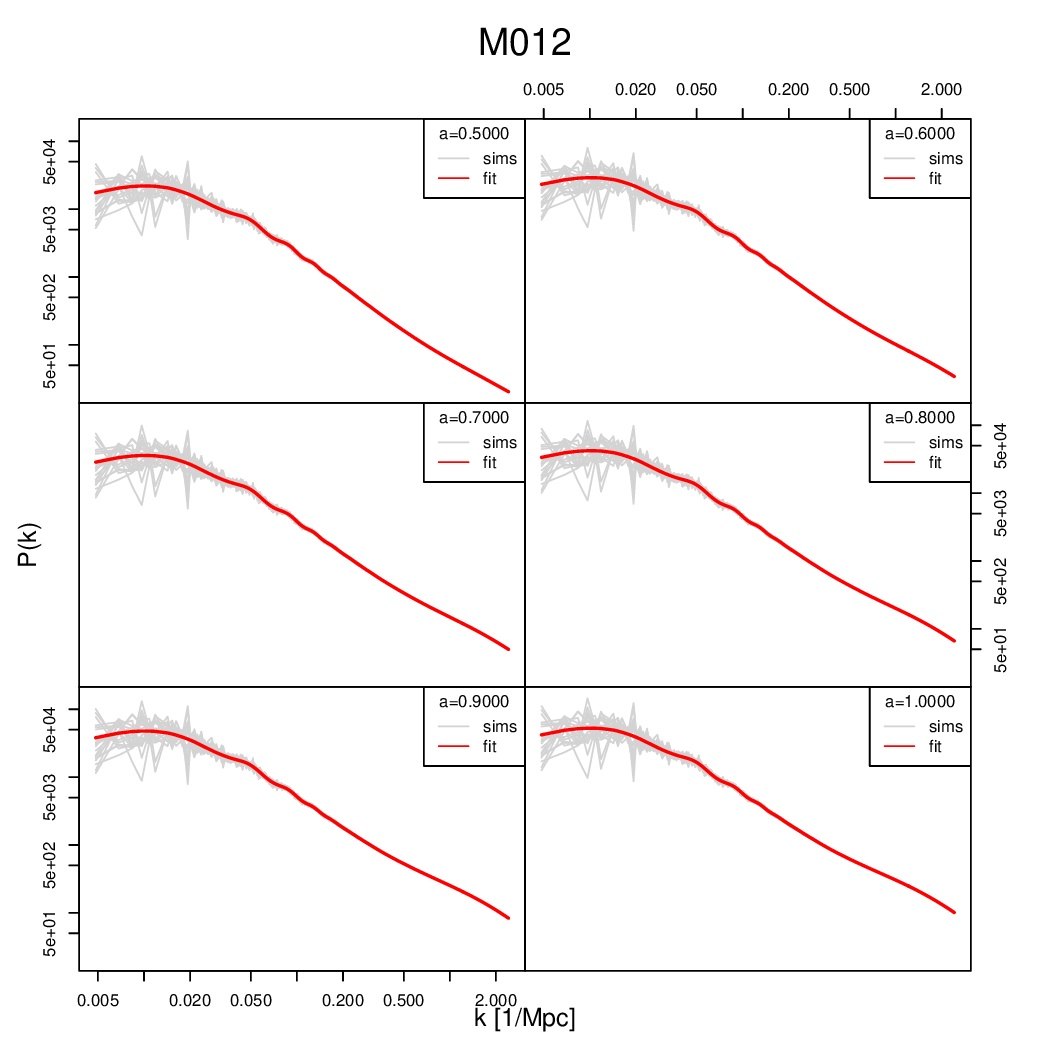}}
\resizebox{2.34in}{!}{\includegraphics{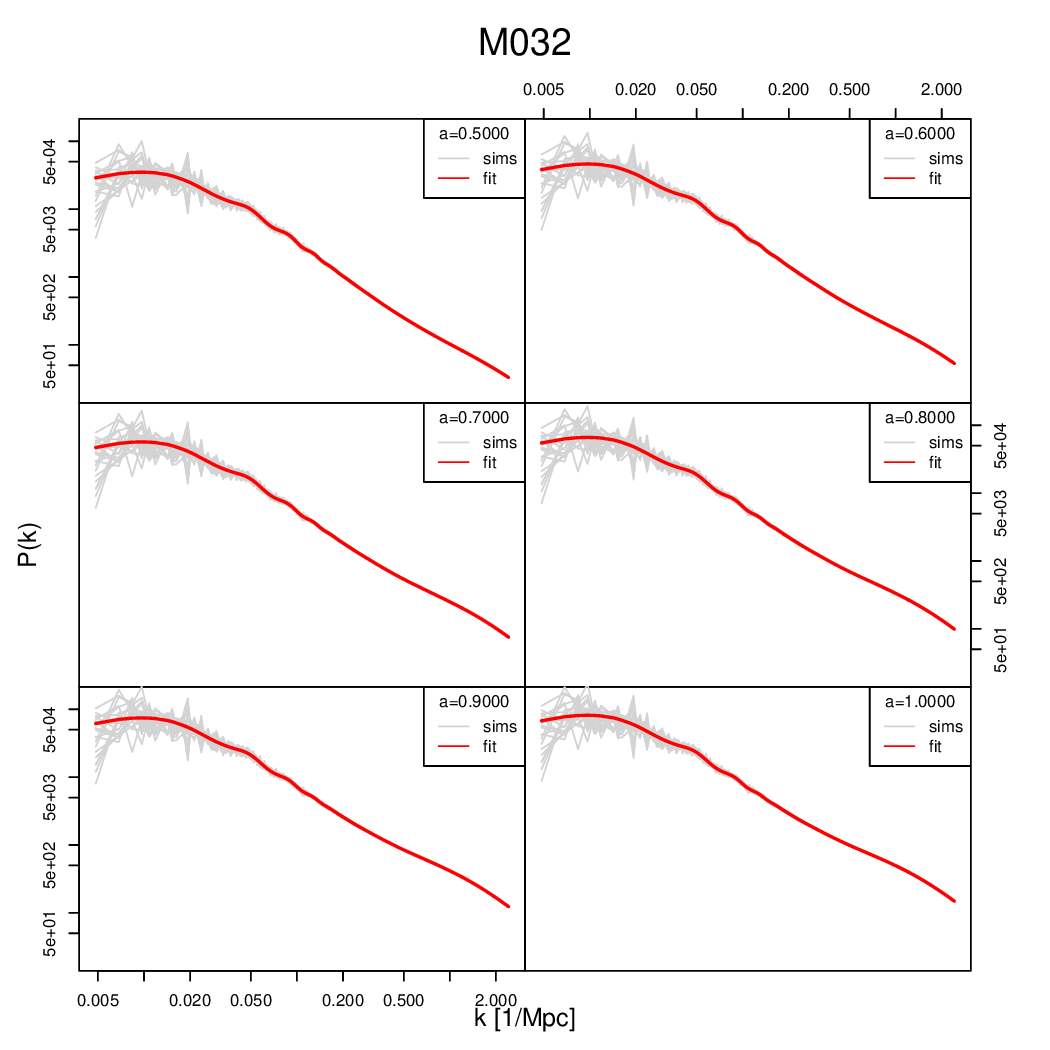}}
\end{center}
\caption{Simulated power spectra and the smooth estimate for each of 
  the three cosmologies at six values for the expansion factor $a$.}
\label{fig:smooths}
\end{figure*}

The next step is to test the matching procedure between theoretical
and simulation results to ensure high accuracy at the perturbation
theory matching point. Since we do not know the exact answer for the
nonlinear power spectrum, we first carry out a test using linear
theory. In this test we use the power spectra from the initial
conditions and show how well we can smooth out the run-to-run scatter.
Knowing the exact answer here will allow us to assess how well the
matching procedure actually works.

For these spectra, the entire estimation procedure is applied to the
initial condition spectra.  The only difference is that fewer grid
points were used for the latent processes $u$ and $v$ to account for
the reduced range of the spectra (we use 70 and 7, respectively).  The
final results are shown in Figure~\ref{fig:lin_pred} for the same set
of three random models as in Figure~\ref{fig:compare}. We verified
that the results hold for all of the remaining models. The upper
panels of each sub-panel show the theoretical linear power spectrum in
black and the prediction from the simulations in red. The vertical
line marks the matching point to linear theory at $k\simeq
0.03$Mpc$^{-1}$. The lower panels show the ratio of the predicted
power spectra to the theoretical power spectra in red. Below the
matching point, the agreement is -- by construction -- perfect. Beyond
the matching point, the smoothed prediction in red is accurate at the
1\% level. This test shows that we can obtain a smooth, high-accuracy
prediction for the power spectrum by combining a suite of realizations
with perturbation theory.

\subsection{The Nonlinear Power Spectrum}
The process convolution procedure is applied to the power spectrum
realizations for each of the 37 cosmologies at six values of
the scale factor $a = 1 / (1+z) \in \{0.5, 0.6, 0.7, 0.8, 0.9, 1.0\}$,
where $z$ is the redshift. The results are shown in
Figure~\ref{fig:smooths}, again for a subset of three models. Although
we have no known truth to use as a comparison in this case, the
resulting estimates continue to fit the simulation realizations very well.  

\begin{figure*}[t]
\begin{center}
\resizebox{2.34in}{!}{\includegraphics{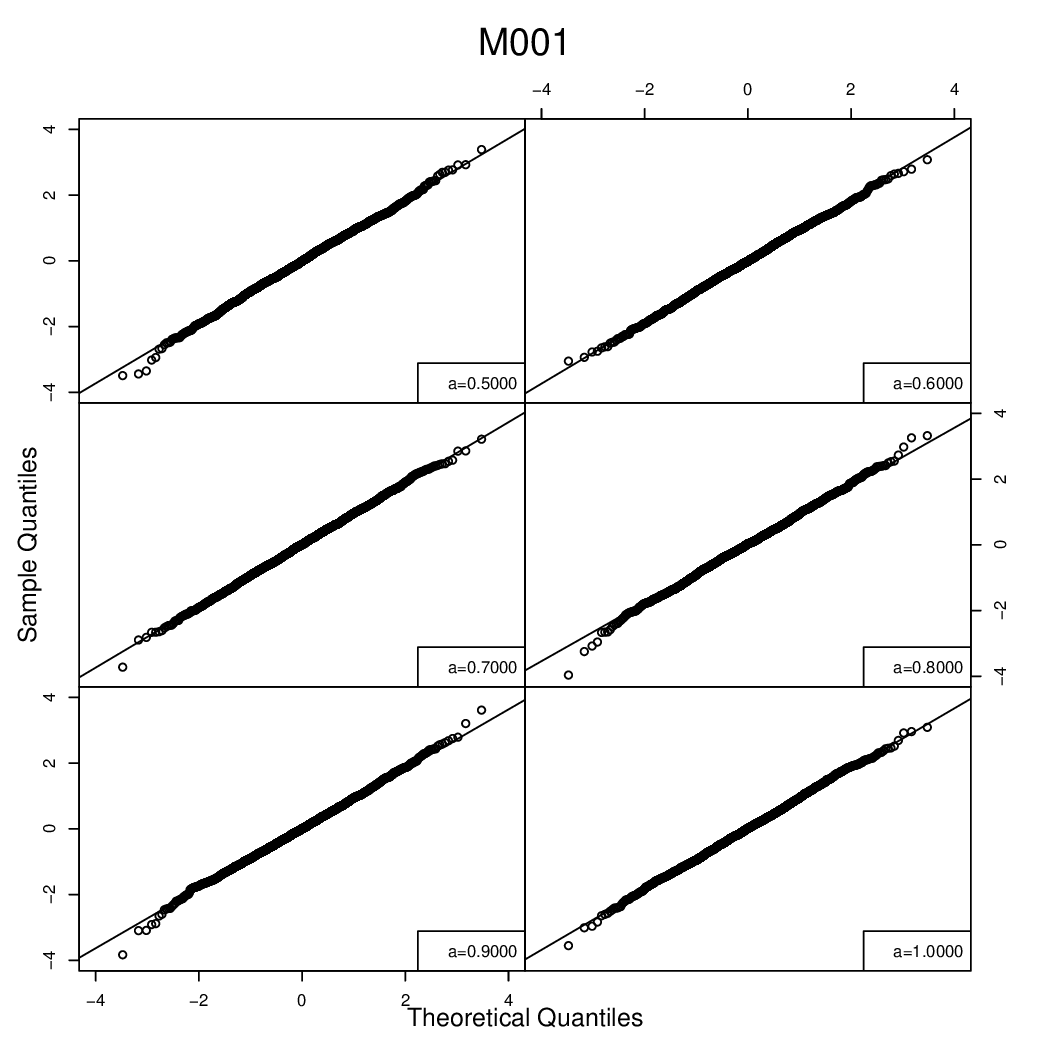}}
\resizebox{2.34in}{!}{\includegraphics{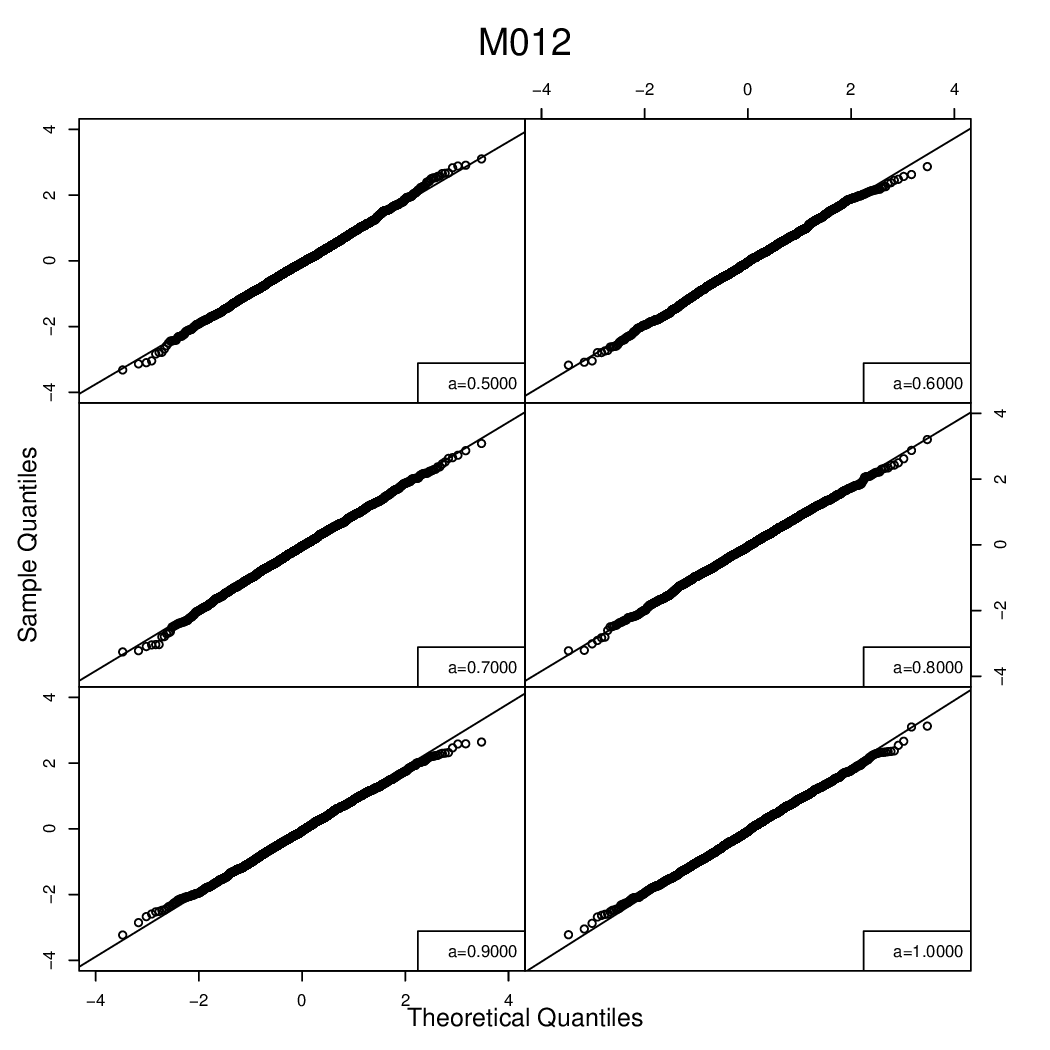}}
\resizebox{2.34in}{!}{\includegraphics{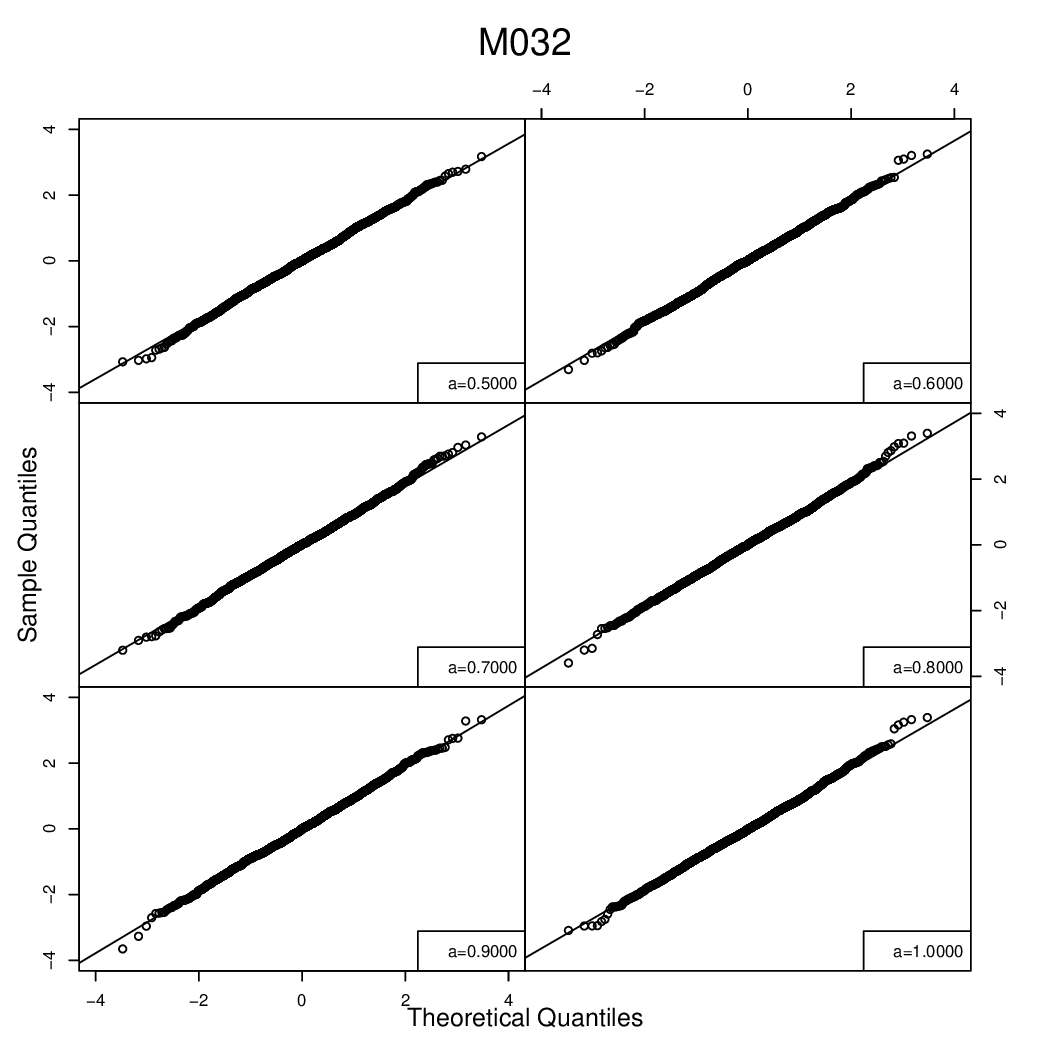}}
\end{center}
\caption{Quantile-quantile plots of the standardized residuals for the G
simulation at the six scale factors for three cosmologies.
Standardized residuals are computed by subtracting the estimated mean
from the simulation and multiplying each value by the square root of
the known precision at its k value.  Our assumptions suggest that the
resulting sample should follow a standard normal distribution with no
dependence on k.  These plots show the sample quantiles of the
standardized residuals (essentially the sorted values) plotted against
the theoretical quantiles for a sample of this size from the standard
normal distribution.  The nearly straight lines indicate little
deviation from our assumptions and suggest that the model fits well.}
\label{fig:qq}
\end{figure*}

\begin{figure*}[t]
\begin{center}
\resizebox{2.34in}{!}{\includegraphics{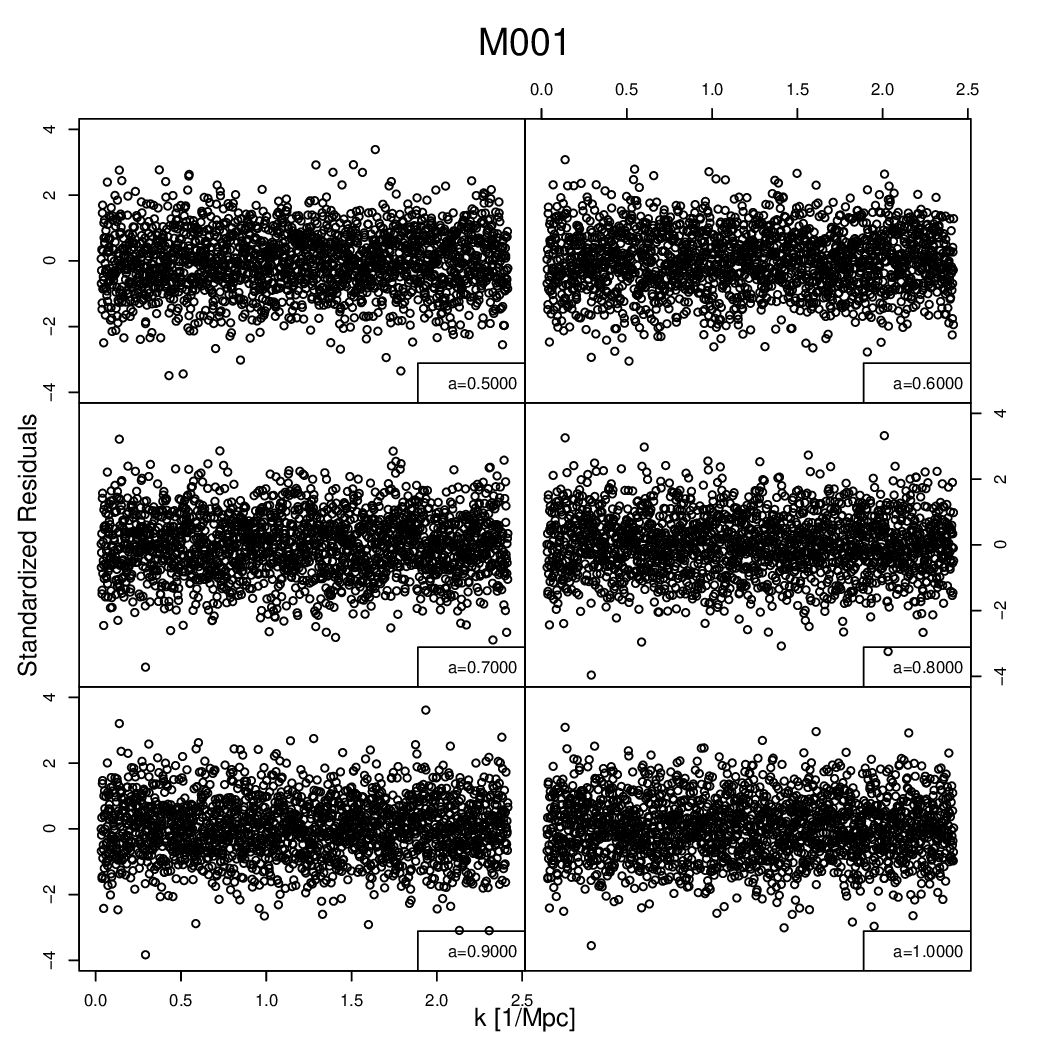}}
\resizebox{2.34in}{!}{\includegraphics{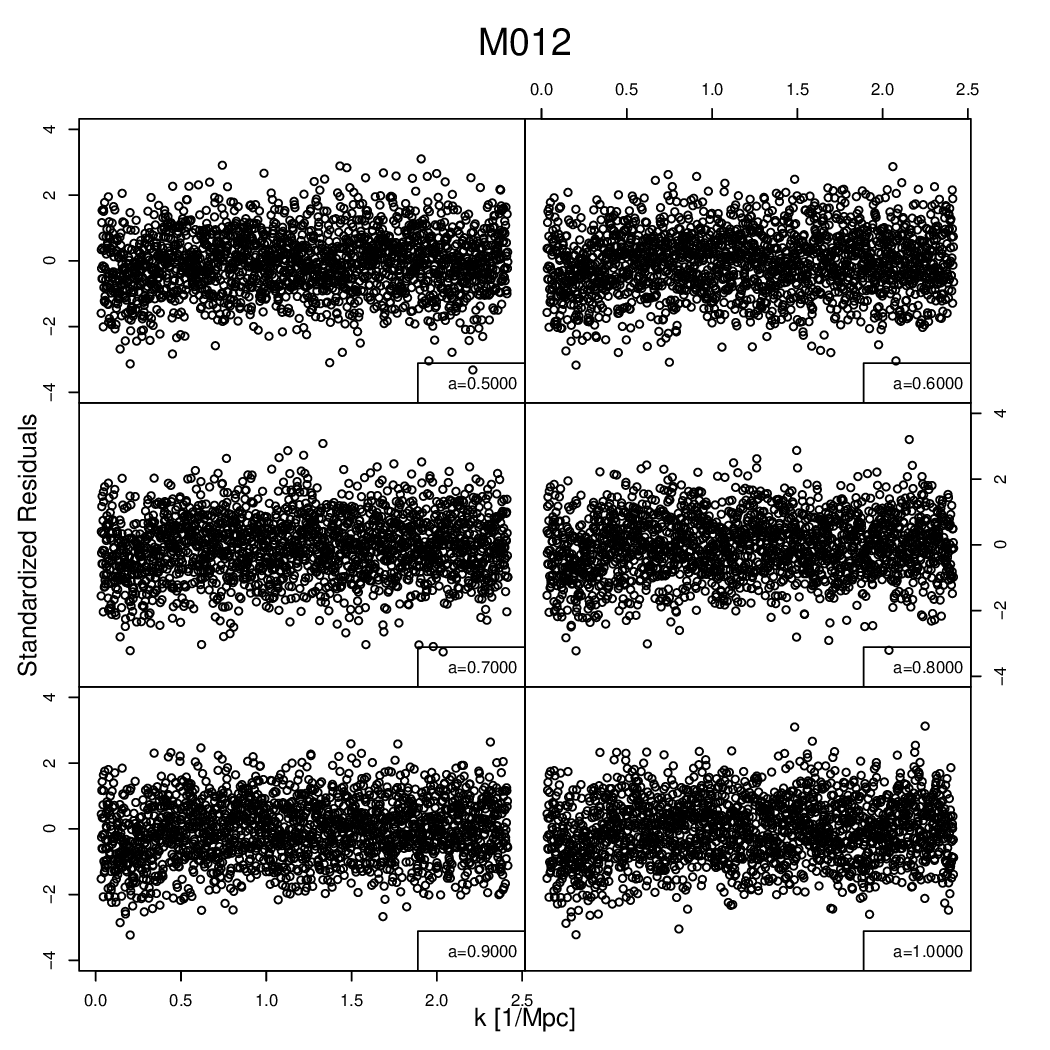}}
\resizebox{2.34in}{!}{\includegraphics{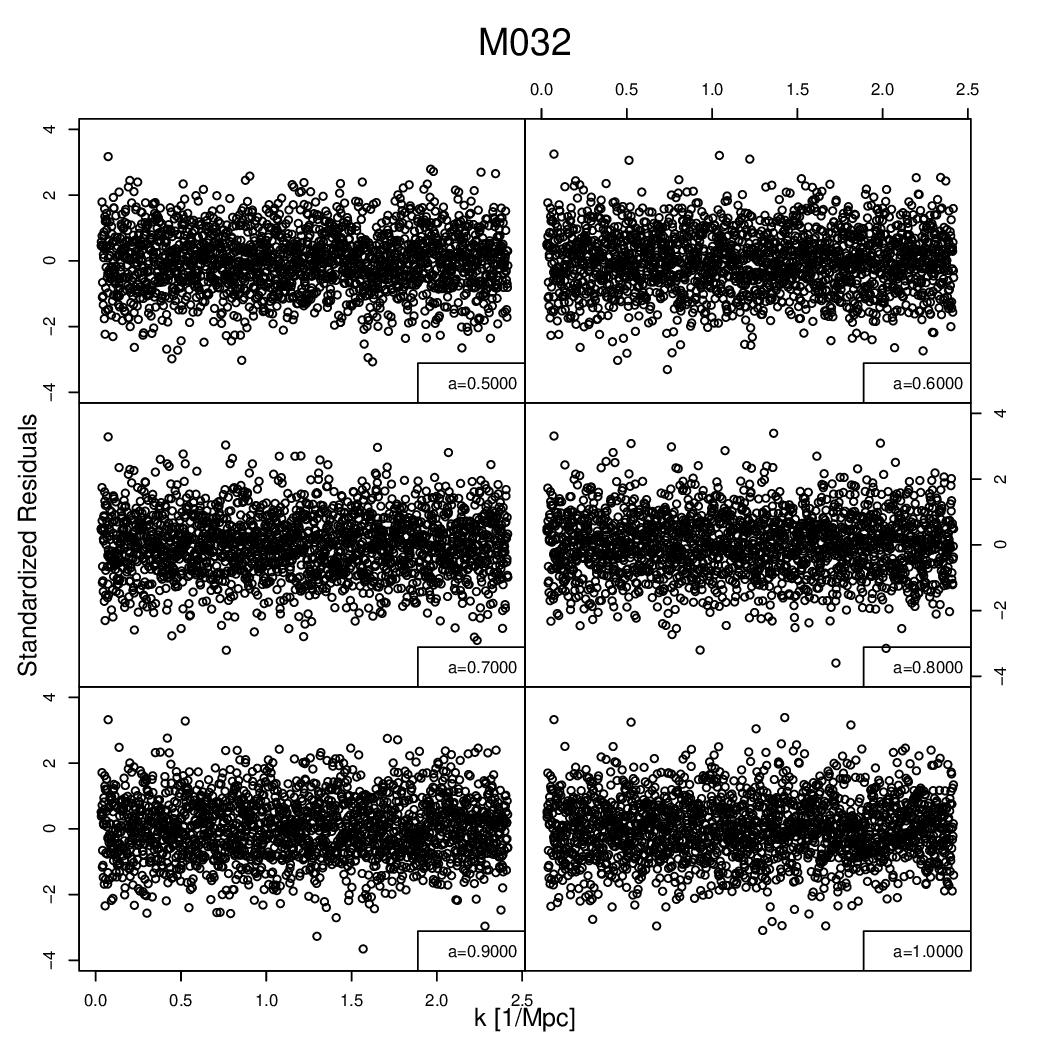}}
\end{center}
\caption{The standardized residuals 
for the G simulation at the six scale factors for three cosmologies
plotted against $k$.  There are no obvious correlations or structure
confirming our assumptions.}
\label{fig:res}
\end{figure*}

In place of comparing with known truth, we can examine some tests of
our modeling assumptions.  We assume that the simulation spectra on
the modified scale are independently and normally distributed about a
smooth mean with a variance that changes with $k$.  Given this, we can
compute standardized residuals in which the estimated smooth mean is
subtracted from the simulations and the result is scaled by the known
standard deviation.  These standardized residuals should look like
\textit{i.i.d.} standard normal variables.  Figure \ref{fig:qq} shows
quantile-quantile plots for the G simulations of three cosmologies at
each of the six scale factors.  Theoretical quantiles from a standard
normal distribution are given on the $x$-axis and the sample quantiles
of the standardized residuals are shown on the $y$-axis.  The nearly
straight line at $45^{\circ}$ indicates an extremely good
distributional fit.  Figure \ref{fig:res} shows the standardized
residuals plotted against $k$ for the same simulations.  This plot
verifies the independence assumption.  Further, there is an almost
complete lack of evidence of any structure in these plots, suggesting
that we would not greatly improve the fit by relaxing the smoothness
assumption.

\begin{figure}[t]
\begin{center}
\resizebox{3in}{!}{\includegraphics{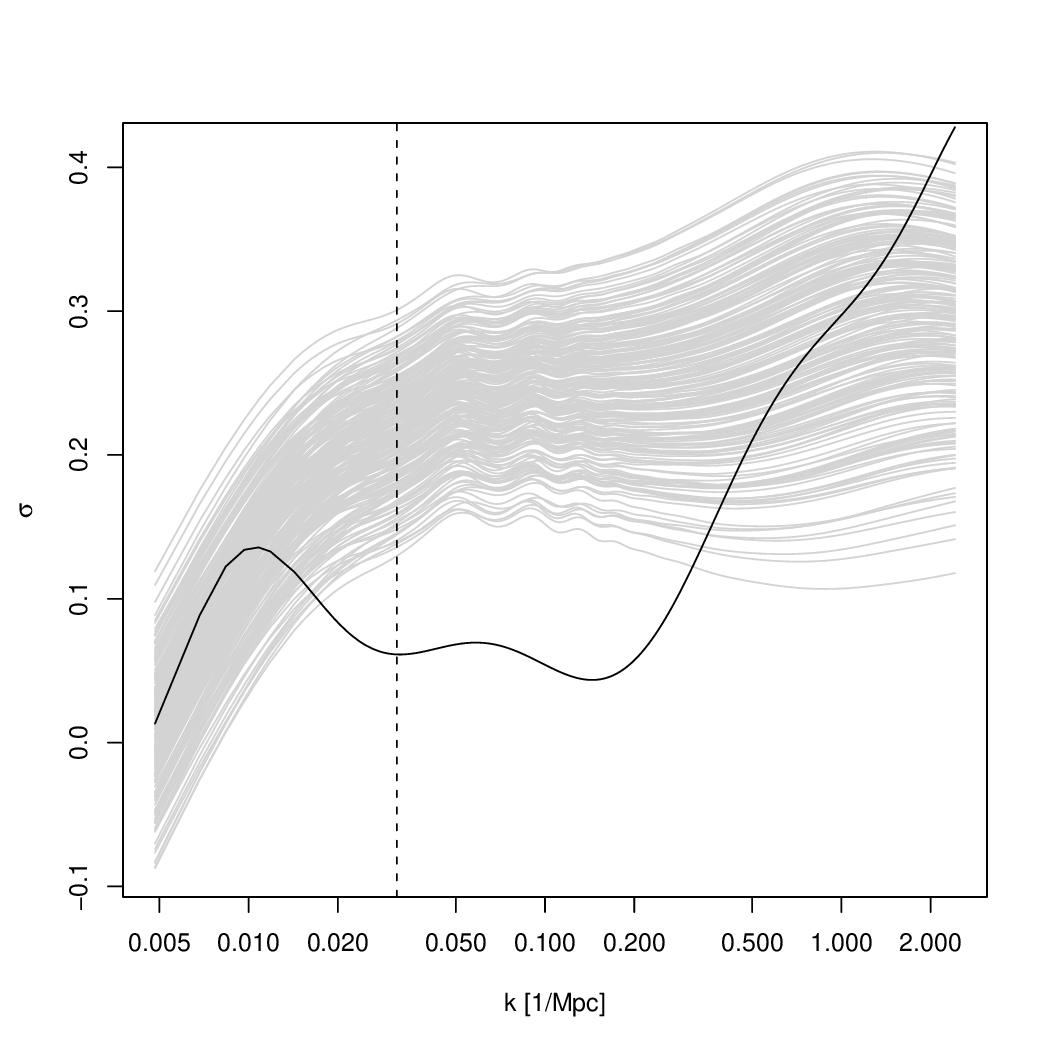}}
\end{center}
\caption{Median MCMC draw for the bandwidth function $\sigma$ (black line).
  Small values on this plot correspond to places where the spectra are
  comparatively less smooth because of the baryon wiggles. In
  addition, all the power spectra are shown (37 models, 6 redshifts
  each) in light gray. The power spectra a scaled and shifted to fit
  the plot. $\sigma$ is low on the baryon acoustic oscillation scale
  to accommodate local wiggles, as expected. The vertical line shows
  the matching point between perturbation theory and the simulation
  outputs.  On the left of this, each replicate is identical (compared
  to the realizations from the different simulation boxes) and smooth,
  therefore the behavior of $\sigma$ has not much information. }
\label{fig:sigma}
\end{figure}

It is also interesting to examine the plot of the kernel width
function as estimated by the MCMC process.  Figure \ref{fig:sigma}
shows the median draw for $\sigma$ as a function of $k$.  As expected,
the kernel width is small in the vicinity of the baryon wiggles.  This
means the values of the latent process $u$ in this region receive
large weights, with the contributions for values further away dropping
off very quickly.  On both ends, the kernel width is large, which
results in a smooth function in these regions.

\section{The Emulator}
\label{sec:emulator}

Having extracted the smooth power spectra from our simulation
suite, we can now build an emulator to predict the nonlinear
matter power spectrum within the priors specified in
Eqns.~(\ref{priors}). We will use only models 1 - 37 for the emulator
construction; model M000 will serve as an independent check of the
emulator accuracy, along with hold-out tests described below.

In order to construct the emulator, we model the 37 power spectra
using an $n_\mathcal{P}$ dimensional basis representation: 
\begin{equation}
\mathcal{P}(k;z;\theta)=
\sum_{i=1}^{n_\mathcal{P}}\phi_i(k;z)w_i(\theta),~~~\theta\in
[0,1]^{n_\mathcal{\theta}}, 
\end{equation}
where the $\phi_i(k;z)$ are the basis functions, the $w_i(\theta)$ are
the corresponding weights, and the $\theta$ represent the cosmological
parameters. The dimensionality $n_\mathcal{P}$ refers to the number of
orthogonal basis vectors
$\{\phi_i(k,z),\dots,\phi_{n_\mathcal{P}}(k,z)\}$. The parameter
$n_\theta$ is the dimensionality of our parameter space -- with 5
cosmological parameters we have $n_\theta=5$.  The power spectrum
$\mathcal{P}(k;z;\theta)$ depends on the wavenumber $k$, the redshift
$z$, and the five cosmological input parameters $\theta$ (note that we
rescale the range of each parameter to $[0,1]$). Examination of the
results indicates that $n_\mathcal{P}=5$ is a good choice for the
number of basis vectors $\phi_i(k;z)$ (that $n_\mathcal{P}=n_\theta$
here is a coincidence). The task is now to (1) construct a suitable
set of orthogonal basis vectors $\phi_i(k;z)$ and (2) model the
weights $w_i(\theta)$). For the first task we use principal
components, and for the second, Gaussian Process (GP) models. Our
choice of GP modeling is based on their success in representing
functions that change smoothly with parameter variation, e.g., the
variation of the power spectrum as a function of cosmological
parameters. Both steps are explained in great detail in Paper II; we
refer the interested reader to that publication. Paper II also
contains various error control tests of the GP-based interpolation
method which we do not repeat here.

Since the details of how to build an emulator are already provided in
Paper II we can immediately turn to our final product: the emulator
itself. To facilitate use of the emulator, we are releasing the fully
trained emulator with this paper.  It provides nonlinear power spectra
at a set of redshifts between $z=0$ and $z=1$ out to
$k=1~h$Mpc$^{-1}$, for any cosmological model specified within the
priors given in Eqns.~(\ref{priors}).

\subsection{Emulator Test}
\begin{figure}[t]
\begin{center}
\resizebox{3in}{!}{\includegraphics{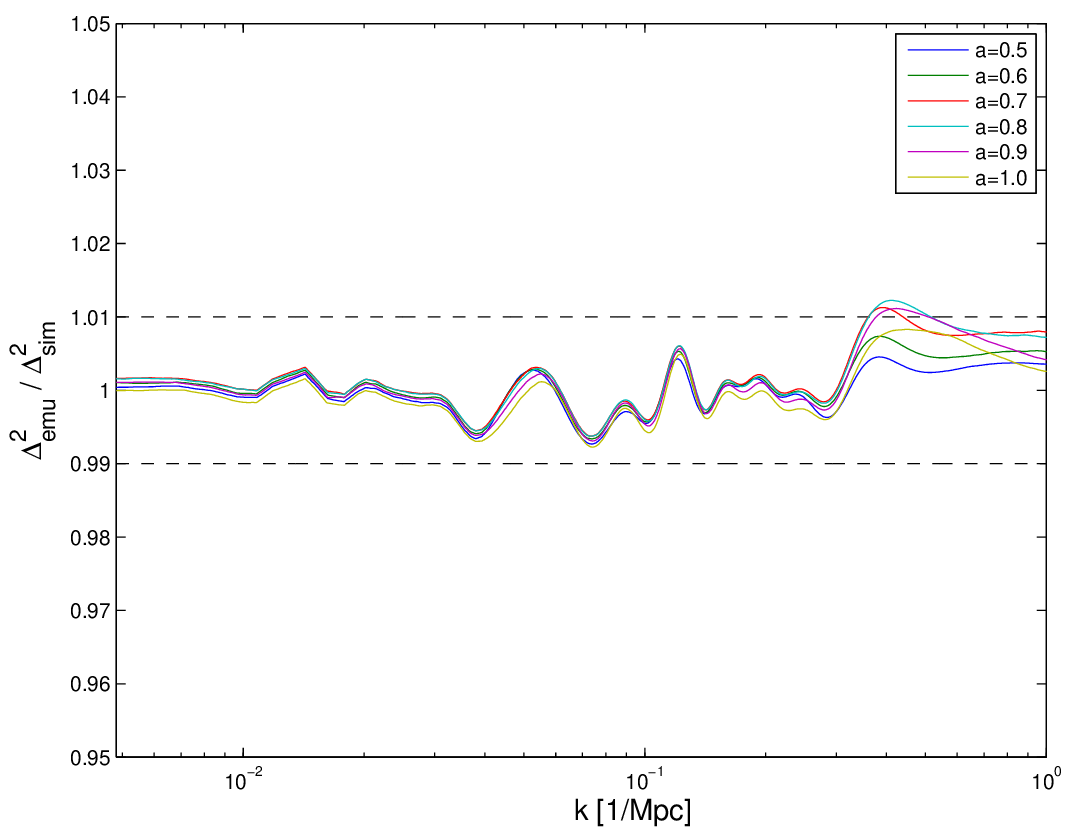}}
\end{center}
\caption{Ratio of the emulator prediction to the smooth simulated
  power spectra for the M000 cosmology at six values of the scale
  factor $a$. The error exceeds 1\% very slightly in only one part of
  the domain for the scale factors $a=0.7, 0.8$, and 0.9.}
\label{fig:M000holdout}
\end{figure}

In order to verify the accuracy of the emulator, we perform two
important checks (other tests on the methodology were carried out in
Paper II). For the first, we compare the simulation results for model
M000 with the emulator prediction. Figure \ref{fig:M000holdout}
shows the results of this true out-of-sample prediction. The plot
shows the ratio of the emulated power spectrum to the actual simulation
for the M000 cosmology at six values of the scale factor $a$. This
cosmology is completely interior in the design and, since M000 was not
used to estimate the smoothing or the emulator, this test provides a
good indication of the actual performance of the emulator within its
parameter bounds. Overall the agreement between the simulations and
the emulator is excellent and for most of the $k$-range well below the
1\% error bound.

\begin{figure}[t]
\begin{center}
\resizebox{3in}{!}{\includegraphics{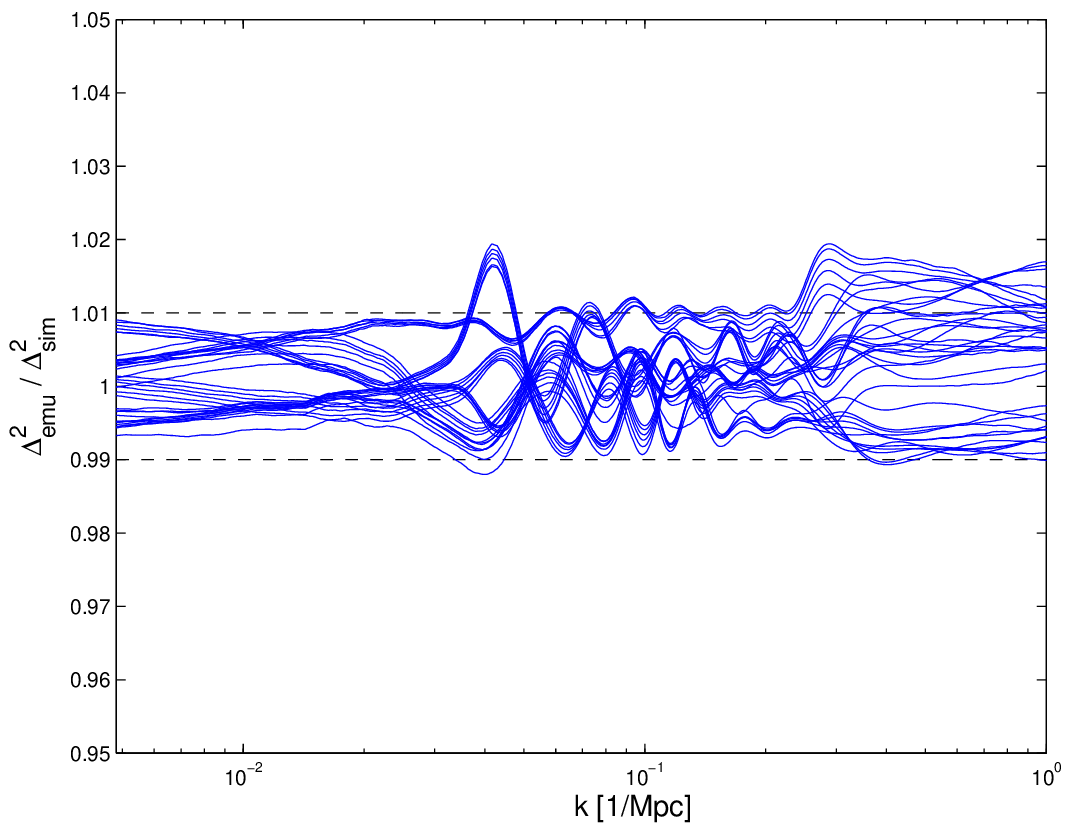}}
\end{center}
\caption{M004, M008, M013, M016, M020, and M026.  For each simulation,
  we show six residuals corresponding to the different values of the
  scale factor $a$.  The errors are on the order of 1\% for the bulk
  of the domain of interest. Considering that the tested emulators are
  built on an incomplete design, this result is remarkably good.}
\label{fig:holdout}
\end{figure}

The second check consists of a sequence of holdout tests. In a holdout
test, the emulator is built from 36 cosmological models and the
emulator prediction can be then compared to the result from the extra
model. The drawback of this test is that if we have only a very small
number of models, each of them is important for capturing some part of the
parameter space and taking it out for building the emulator degrades
the emulator precision. In order to keep this problem to a minimum, we
only perform holdout tests for models which are interior simulations,
meaning that none of the five parameters are near the extreme limits
of the chosen prior range. There are six such models: M004, M008,
M013, M016, M020, and M026. The ratio of the emulated prediction to
the actual simulation for these models is shown in
Figure~\ref{fig:holdout}. For each cosmology, there are six ratios for
each value of the scale factor. The lines are quite well behaved with
errors largely within the 1\% bounds.

\section{Lessons Learned and Future Challenges}
\label{sec:lessons}

The advent of precision cosmology and the prospect of very large
surveys such as LSST and JDEM poses an enormous challenge to the
theory community in the field of large scale structure predictions.
Future progress not only requires {\em very accurate} predictions but
also the ability to produce predictions for different cosmological
models {\em very fast}. The aim of the Coyote Universe project was to
take a first step in attacking this problem, focusing on the matter
power spectrum at intermediate scales out to $k\simeq
1$~$h$Mpc$^{-1}$. While predicting the matter power spectrum on these
scales at high accuracy may appear to be a moderately difficult task,
it turned out to be a technical and computational challenge in several
respects. It is generally hard to imagine problems and pitfalls in
advance, which is why fully working through an example is so helpful.
Since the community has to follow a similar path in the future to
create predictive capabilities for different cosmological probes, we
summarize here some of the lessons learned during this project.

The major differences between a project like this (including
all three papers of the series) and previous numerical studies of
large scale structure probes are:

(1) {\em Computational and Storage Capacity:} The Coyote Universe
simulation suite encompasses roughly 60 TB of data. The computational
cost is of the order of a million CPU-hours and including waiting
times in submission queues, downtimes of the machine, and so on,
carrying out these simulations took roughly six months. The simulation
size of one billion particles in a Gpc$^3$ volume was barely enough to
resolve the scales of interest and for future work will certainly not
be sufficient. Such simulations will need larger volumes and better
mass resolution. For example, to resolve a 10$^{12}$M$_\odot$ halo
with 100 particles in a (3~Gpc)$^3$ volume, we would need 300 billion
particles. While supercomputers will get faster and larger in the
future, generating many simulations at the edge of machine
capabilities will always be a challenge. In addition, archiving the
outputs of such simulations will become very expensive in terms of
storage. From the Coyote Universe runs we stored 11 time snapshots
plus the initial conditions (particle positions and velocities and
halo information) leading to 250~GB of data per run. For the 300
billion particle run this would increase to 75~TB. Only very few
places worldwide would be able to manage the resulting large
databases.

(2) {\em Simulation Infrastructure:} Running a very large number of
simulations makes it necessary to integrate the major parts of the
analysis steps into the simulation code and to automate as much of the
mechanics of running the code (submission, restarts) as possible. For
the Coyote Universe project we developed several scripts to generate
the input files of the codes, to structure the directories where
different runs are performed in, and to submit the simulations to the
computing queue system. For future efforts of this kind the adoption
and development of dedicated workflow capabilities for these tasks
must be considered. The number of tasks to be carried out will become
too large to keep track of without such tools. In addition, since
large projects will require extensive collaborations, software tools
will make it easier to work in a team environment since each
collaborator will have information about previous tasks and
results. An example of such a tool for cosmological simulation
analysis and visualization is given in \cite{anderson}.

We carried out the data analysis after the runs were finished. For
very large simulations this is not very practical, and on-the-fly
analysis tools are required to minimize read and write times and
failures. This in turn requires that the code infrastructure be
tailored to the problem under consideration.

(3) {\em Serving the Data:} Clearly, large simulation efforts cannot
be carried out by a few individuals, and require possibly
community wide coordination. The simulation data will be valuable for
many different projects. It is therefore necessary to make the data
from such simulation efforts publicly available and serve them in a
way that new science can be extracted from different groups of
simulations. Transferring large amounts of data is difficult because
of limitations in communication bandwidth and also because of the
large storage requirements. It would therefore be desirable to have
computational resources dedicated to the database. In such a
situation, researchers would be able to run their analysis codes on
machines with direct access to the database and perform queries on the data
easily. We are planning to make the Coyote Universe database available
in the future and use it as a manageable testbed for such services.

(4) {\em Communication with other Communities:} The complexity of the
analysis task makes it necessary to efficiently collaborate and
communicate with other communities, for example statisticians,
computer scientists, and applied mathematicians. Many tools that will
be essential for precision cosmology in the future have already been
invented -- the task is to find them and use them in the best way
possible.

\section{Conclusions}
\label{sec:conclusions}

This paper is the last of the Coyote Universe series of publications.
Paper~I was concerned with demonstrating that percent level accuracy
in the (gravity only) nonlinear power spectrum could be attained out
to $k\simeq 1$ Mpc$^{-1}$. Paper~II showed that with only a relatively
small number of simulations, interpolation across a high-dimensional
space was possible at close to the same level of accuracy as that
attained in the individual runs. Paper~III takes this work to the
final conclusion: Based on almost 1000 simulations spanning 38 $w$CDM
cosmologies, we present a fast and very accurate prediction scheme --
an emulator -- for the nonlinear matter power spectrum. The emulator
is accurate at the percent level, improving over commonly used fitting
functions by almost an order of magnitude.

The emulator construction -- as explained in Paper~II -- is based on
GP modeling. In order to carry this out, a major challenge is to
produce a smooth power spectrum from a finite set
of simulations for each cosmology. In order to minimize run-to-run scatter on very large
scales, we performed several medium resolution simulations and matched
these at sufficiently low $k$ to perturbation theory. On quasi-linear
scales we used medium resolution simulations and matched those to high
resolution runs at small spatial scales. Matching the different
resolution runs and perturbation theory accurately was carried out
using process convolution. This technique allowed us to construct a
smooth power spectrum for each cosmological model. The results were
then used to construct the emulator via GP modeling as described in
detail in Paper~II.

We are releasing the power spectrum emulator as a C code,
which allows the user to specify a cosmology within our priors and
returns the power spectrum at six different redshifts between $z=0$
and $z=1$ out to $k\simeq 1$ Mpc$^{-1}$. These power spectra can now
be used for further analysis of cosmological data. We are planning to
extend the emulator in the near future to a larger $k$-range and will
provide a smooth interpolation between results from different
redshifts.

A major challenge will be to ensure that a certain level of accuracy
is reached with the simulations when going to small scales. Besides
being computationally very expensive (high force and mass resolution
being required) the physics at smaller scales is far more complicated.
The inclusion of gasdynamics and feedback effects (along with other
physics) is far from being straightforward. The impossibility of
carrying out a direct simulation effort is more or less certain; a
good number of phenomenological/subgrid modeling parameters will be
required. As simulation complexity and the number of modeling and
cosmological parameters increases, it becomes even more important to
develop efficient and controlled sampling schemes as described in
Paper~II, so that data can be used to determine both cosmological and
modeling parameters (self-calibration).

With the series of three Coyote Universe papers we have demonstrated
that it is possible to extract cosmological statistics such as the
power spectrum at high accuracy and that one can build an accurate
prediction scheme based on a limited set of simulations. This line of
work will be important for interpreting results of future cosmological
surveys. It will also have to be extended in several ways: (1) the
cosmological model space has to be opened up; (2) we have to ensure
high accuracy at scales smaller than those considered here; (3) we
have to include more physics in order to capture those small scales
correctly; (4) we have to include different cosmological probes, e.g.,
the cluster mass function and the shear power spectrum, to be
able to build a complete framework for analyzing future survey data.
We have shown here that such a program can in principle be
established, though it will demand a large concerted effort between
different communities.

\acknowledgements

A special acknowledgment is due to supercomputing time awarded to us
under the LANL Institutional Computing Initiative. Part of this
research was supported by the DOE under contract W-7405-ENG-36 and by
a DOE HEP Dark Energy R\&D award. S.H., K.H., D.H., E.L., and C.W.
acknowledge support from the LDRD program at Los Alamos National
Laboratory. K.H. was supported in part by NASA. M.W. was supported in
part by NASA and the DOE. We would like to thank Dragan Huterer,
Nikhil Padmanabhan, Adrian Pope, and Michael Schneider for useful
discussions. We thank Volker Springel for making the $N$-body code
{\sc GADGET-2} publicly available.


\begin{thebibliography}{99.}

\bibitem[{{Albrecht et al.}(2006)}]{DETF}
Albrecht,~A. et al. 2006, arXiv:astro-ph/0609591.

\bibitem[{{Albrecht et al.}(2009)}]{FOM}
Albrecht,~A. et al. 2009, arXiv:0901.0721.

\bibitem[{{Anderson et al.}(2008)}]{anderson} 
Anderson, E., Silva, C., Ahrens, J., Heitmann, K., \& Habib,~S. 2008,
Computing in Science and Engineering, 10, 30 

\bibitem[{{Carlson et al.}(2009)}]{carlson09}
Carlson, J.W., White, M., \& Padmanabhan, N. 2009,  Phys. Rev. D, 80,
043531 

\bibitem[{{Chib \& Greenberg}(1995)}]{ChibGreenberg1995} 
Chib,~S. and Greenberg,~E. 1995, The American Statistician, 49, 327  

\bibitem[{{Goroff et al.}(1986)}]{Gor86}
Goroff,~M.H., Grinstein,~B., Rey,~S.-J., \& Wise,~M.B. 1986, ApJ,
311, 6  

\bibitem[{{Habib et al.}(2007)}]{HHHNW}
Habib,~S., Heitmann,~K., Higdon,~D., Nakhleh,~C., \& Williams,~B.
2007, Phys. Rev. D, 76, 083503

\bibitem[{{Heitmann et al.}(2006)}]{HHHN}
Heitmann,~K., Higdon,~D., Nakhleh,~C., \& Habib,~S. 2006, ApJ 646, L1 

\bibitem[{{Heitmann et al.}(2009a)}]{CoyoteI}
Heitmann,~K., White,~M., Wagner,~C., Habib,~S., \& Higdon,~D., ApJ
(submitted) [arXiv:0812.1052] (Paper I)

\bibitem[{{Heitmann et al.}(2009b)}]{CoyoteII}
Heitmann,~K., Higdon,~D., White,~M., Habib,~S., Williams,~B., \&
Wagner,~C. 2009, ApJ, 705, 156 (Paper II) 

\bibitem[{{Higdon}(2002)}]{Higdon2002}
Higdon,~D. 2002, Space and Space-Time Modeling Using Process Convolutions
in {\it Quantitative Methods for Current Environmental Issues}  (Springer)

\bibitem[{{Jain \& Bertschinger}(1994)}]{JaiBer94}
Jain,~B. and Bertschinger,~E. 1994, ApJ, 431, 495

\bibitem[{{Juszkiewicz}(1981)}]{Jus81}
Juszkiewicz,~R. 1981, MNRAS, 197, 931

\bibitem[{{Komatsu et al.}(2008)}]{WMAP5}
Komatsu,~E. et al. 2009, ApJS, 180, 330

\bibitem[{{Makino et al.}(1992)}]{Mak92}
Makino,~N., Sasaki,~M., \& Suto,~Y. 1992, PRD 46, 585

\bibitem[{{Matsubara}(2008)}]{mats08}
Matsubara,~T. 2008, Phys. Rev. D, 77, 063530

\bibitem[{{Meiksin, White \& Peacock}(1999)}]{MWP99}
Meiksin,~A., White,~M., \& Peacock,~J.A. 1999, MNRAS, 304, 851

\bibitem[{{Peacock \& Dodds}(1996)}]{PD96}
Peacock,~J.A. and Dodds,~S.J. 1996, MNRAS, 280, L19

\bibitem[{{Peebles}(1980)}]{Pee80}
Peebles,~P.J.E. 1980, {\em The Large-Scale structure of the Universe},
Princeton University Press (Princeton).

\bibitem[{{Perlmutter et al.}(1999)}]{perlmutter}
Perlmutter,~S. et al. 1999, ApJ, 517, 565

\bibitem[{{Riess et al.}(1998)}]{riess} 
Riess,~A.G, et al. 1998, AJ, 116, 1009

\bibitem[{{Schneider et al.}(2008)}]{SKHHHN}
Schneider,~M., Knox,~L., Habib,~S., Heitmann,~K., Higdon,~D., \&
Nakhleh,~C. 2008, Phys. Rev. D, 78, 063529 

\bibitem[{{Smith et al.}(2003)}]{Smi03} 
Smith,~R.E., et al.  [The Virgo Consortium Collaboration] 2003, MNRAS, 341, 1311

\bibitem[{{Springel}(2005)}]{Spr05}
Springel,~V.  2005, MNRAS, 364, 1105

\bibitem[{{Takahashi et al.}(2008)}]{taka08}
Takahashi,~R. et al. 2008, MNRAS, 389, 1675

\bibitem[{{Vishniac}(1983)}]{Vis83}
Vishniac,~E. 1983, MNRAS, 203, 345


\end{thebibliography}
\end{document}